\documentclass[aps,nofootinbib,twocolumn,prd,eqsecnum,showpacs,showkeys,preprintnumbers]{revtex4-1}
\usepackage[caption=false]{subfig}
\usepackage{graphicx}
\usepackage{amsmath}
\usepackage{amsfonts}
\usepackage{amssymb}
\usepackage{color}
\usepackage{bm}
\usepackage{mathrsfs}
\usepackage{epstopdf}
\usepackage{url}
\usepackage{footnote}
\usepackage{textcomp}

\makeatletter
\newcommand*{\rom}[1]{\expandafter\@slowromancap\romannumeral #1@}
\makeatother

\begin{document}

\title{Eddington-Born-Infeld cosmology: a cosmographic approach, a tale of doomsdays and the fate of bound structures}
\author{Mariam Bouhmadi-L\'{o}pez$^{1,2}$}
\email{mariam.bouhmadi@ehu.es}
\author{Che-Yu Chen $^{3,5}$}
\email{b97202056@ntu.edu.tw}
\author{Pisin Chen $^{3,4,5,6}$}
\email{pisinchen@phys.ntu.edu.tw}
\date{\today}

\affiliation{
${}^1$Department of Theoretical Physics, University of the Basque Country
UPV/EHU, P.O. Box 644, 48080 Bilbao, Spain\\
${}^2$IKERBASQUE, Basque Foundation for Science, 48011, Bilbao, Spain\\
${}^3$Department of Physics, National Taiwan University, Taipei, Taiwan 10617\\
${}^4$LeCosPA, National Taiwan University, Taipei, Taiwan 10617\\
${}^5$Graduate Institute of Astrophysics, National Taiwan University, Taipei, Taiwan 10617\\
${}^6$Kavli Institute for Particle Astrophysics and Cosmology, SLAC National Accelerator Laboratory, Stanford University, Stanford, CA 94305, U.S.A.
}

\begin{abstract}
The Eddington-inspired-Born-Infeld scenario (EiBI) 
can prevent the Big Bang singularity 
for a matter content whose equation of state is constant and positive. In a recent paper \cite{Bouhmadi-Lopez:2013lha} we showed that, on the contrary, it is
impossible to smooth a big rip in the EiBI setup. In fact the situations are still different for other
singularities. In this paper we show that a big freeze
singularity in GR can in some cases be smoothed to a sudden or a type-IV singularity under the EiBI scenario. Similarly, a sudden or a type-IV singularity
in GR can be replaced in some regions of the parameter space by a type-IV singularity or a loitering behaviour, respectively, in the
EiBI framework. Furthermore, we find that the auxiliary metric related to the physical connection usually has
a smoother behaviour than that based on the physical metric. In addition, we show that bound structures close
to a big rip or a little rip will be destroyed before the advent of the singularity and will remain
bound close to a sudden, big freeze or type-IV singularity. We then constrain the model
following a cosmographic approach, which is well-known to be model-independent, for
a given Friedmann-Lema\^itre-Robertson-Walker geometry. It turns out that among the various past
or present singularities, the cosmographic analysis can pick up the physical
region that determines the occurrence of a type-IV singularity or a loitering effect in the past. Moreover,
to determine which of the future singularities or doomsdays is more probable, observational constraints on
higher order cosmographic parameters is required.
\end{abstract}


\maketitle

\section{Introduction}

With no doubt general relativity (GR) is an extremely successful theory about to become centenary \cite{gravitation}. Nevertheless, it is expected to break down at some point at very high energies where quantum effects can become important, for example in the past evolution of the Universe where GR predicts a big bang singularity \cite{largescale}. This is one of the motivations for looking for possible extension of GR. Moreover, it is hoped that modified theories of GR, while preserving the great achievements of GR, would shed some light over the unknown fundamental nature of dark energy or whatsoever stuff that drives the present accelerating expansion of the Universe (see Ref.~\cite{reviewmodifiedGR} and references therein), said in other words: What is the “hand that started recently to rock the cradle”? 

Indeed, several observations, ranging from type Ia Supernovae (SNeIa) \cite{SNa} (which brought the first evidence) to the cosmic microwave background (CMB) \cite{CMB}, the baryon acoustic oscillations (BAO) \cite{BAO}, gamma ray bursts (GRB) \cite{Dainotti:2013cta}  and measures of the Hubble parameter at different redshifts \cite{Hz} among others,  showed that the Universe has entered in the recent past a  state of acceleration if homogeneity and isotropy is assumed on its largest scale. Actually, observations show that such an accelerating state is fuelled by an \textit{effective matter} whose equation of state is pretty much similar to that of a cosmological constant but which could as well deviate from it by leaving room for quintessence and phantom behaviours, the latter being known to induce future singularities (see Ref.~\cite{Kamenshchik:2013naa} and references therein). Therefore, it is of interest to formulate consistent modified theories of gravity that could appease the cosmological singularities and could shed some light over the late-time acceleration of the Universe. Of course, an alternative way to deal with dark energy singularities is to invoke a quantum approach as done on Ref.~\cite{BouhmadiLopez:2004mp}.

A very interesting theory at this regard has been reformulated recently: the Eddington-inspired-Born-Infeld theory (EiBI) \cite{Vollick:2003qp,Vollick:2005gc,Banados:2010ix}, as its name indicates, is based on the gravitational theory proposed by Eddington \cite{Eddington} with an action similar to that of the non-linear electrodynamics of Born and Infeld \cite{Born:1934gh}. Such an EiBI theory is formulated on the Palatini approach, i.e., the connection that appears in the action is {\it not} the Levi-Civita connection of the metric in the theory. For a metric approach to the EiBI theory see Ref.~\cite{Deser:1998rj}. Like Eddington theory \cite{Eddington}, EiBI theory is equivalent to GR in vacuum, however, it differs from it in the presence of matter. Indeed while GR cannot avoid the Big Bang singularity for a universe filled with matter with a constant and a positive equation of state (with flat and hyperbolic spatial section), the EiBI setup does as shown in \cite{Banados:2010ix,Scargill:2012kg}. The EiBI scenario was as well proposed as an alternative to the inflationary paradigm \cite{Avelino:2012ue} through a bounce induced by an evolving equation of state fed by a massive scalar field. This model comes with the bonus of overcoming the tensor instability previously found in the EiBI model in Ref.~\cite{EscamillaRivera:2012vz} (see also \cite{Yang:2013hsa} for an analysis of the scalar and vectorial perturbations for a radiation dominated universe and the studies on the large scale structure formation in Ref.~\cite{Du:2014jka}). Black hole solutions with charged particles and the strong gravitational lensing within the EiBI theory are studied in Ref.~\cite{Wei:2014dka}. Besides, the fulfilment of the energy conditions in the EiBI theory was studied in Ref.~\cite{Delsate:2012ky} and a sufficient condition for singularity avoidance under the fulfilment of the null energy condition was obtained. Additionally, it was shown that the gravitational collapse of non-interacting particles does not lead to singular states in the Newtonian limit \cite{Pani:2012qb}.  Furthermore, the parameter characterizing the theory has been constrained using solar models \cite{Casanellas:2011kf}, neutron stars \cite{Avelino:2012ge}, and nuclear physics \cite{Avelino:2012qe}. Very recently, neutron stars and wormhole solutions on the EiBI theory were analysed in \cite{Harko:2013wka,Harko:2013aya,Sham:2013cya}. Especially in \cite{Sham:2013cya}, the authors showed that the universal relations of the f-mode oscillation \cite{Lau:2009bu}, which is the fundamental mode of the pulsation modes in the neutron stars, and the I-Love-Q relations \cite{Yagi:2013bca} ,which refers to the relation among the moment of inertia, tidal Love numbers (which are parameters measuring the rigidity of a planetary body and the susceptibility of its shape to change in response to a tidal potential) and the quadrupole moment of the neutron stars, found in GR are also valid in the EiBI theory. A theory which combines the EiBI action and the f(R) action is also analysed in Refs.~\cite{Makarenko:2014lxa,Makarenko:2014nca} (see also Ref.~\cite{Odintsov:2014yaa}) . A drawback of this theory is that it shares some pathologies with Palatini f(R) gravity such as curvature singularities at the surface of polytropic stars \cite{Pani:2012qd} (see also \cite{Noller:2013yja}).

We showed recently that despite the Big Bang avoidance in the EiBI setup, the Big Rip \cite{Starobinsky:1999yw,Caldwell:2003vq,Caldwell:1999ew,Carroll:2003st,Chimento:2003qy,Dabrowski:2003jm,GonzalezDiaz:2003rf,GonzalezDiaz:2004vq} is unavoidable in the EiBI phantom model \cite{Bouhmadi-Lopez:2013lha}. In this paper, we will assume an EiBI model and we will carry a thorough analysis of the possible avoidance of the other dark energy related singularities, known as: (i) Sudden, Type II, Big Brake or \textit{Big D\'emarrage} singularity \cite{BouhmadiLopez:2007qb,Barrow:2004xh,Gorini:2003wa,Nojiri:2005sx}, (ii) Type III or Big Breeze singularity \cite{BouhmadiLopez:2007qb,BouhmadiLopez:2006fu,Nojiri:2005sx,Nojiri:2004pf,Nojiri:2005sr},
and (iii) Type IV singularity \cite{BouhmadiLopez:2007qb,Nojiri:2005sx,Nojiri:2008fk,Bamba:2008ut}. Those singularities can show up in GR when a Friedmann-Lema\^itre-Robertson-Walker (FLRW) universe is filled with a Generalized Chaplygin gas (GCG) \cite{BouhmadiLopez:2007qb} (more precisely, a phantom Generalised Chaplygin gas, or pGCG for short) which has a rather chameleonic behaviour despite its simple equation of state \cite{BouhmadiLopez:2007qb,BouhmadiLopez:2006fu}. Indeed, the GCG can unify the role of dark matter and dark energy \cite{Bento:2002ps,Kamenshchik:2001cp} (for a recent update on the subject see Ref.~\cite{Wang:2013qy}),  avoid the Big Rip singularity \cite{BouhmadiLopez:2004me}, describe some primitive epoch of the Universe  \cite{BouhmadiLopez:2009hv} and alleviate the observed low quadruple of the CMB \cite{BouhmadiLopez:2012by}. We will complete our analyses by considering as well the possible avoidance of the Little Rip event \cite{Ruzmaikina1970} on the above mentioned setup.

In the EiBI theory, there are two metrics, the first one $g_{\mu\nu}$ appears in the action and couples to matter, the second one is the auxiliary one which is compatible with the connection $\Gamma$ \cite{Banados:2010ix}. The two metrics reduce to the original one in GR when the curvature term is small. Therefore, we will analyse the singularity avoidance with respect to both metrics. Furthermore, we will use the geodesic equations compatible with both metrics to study the behaviour of the physical radius of a Newtonian bounded system near the singularities. For an exhaustive analysis of the geodesics close to the dark energy related singularities in GR see Refs.~\cite{Nesseris:2004uj,Faraoni:2007es}. As a result, we find that the asymptotic behaviour of $g_{\mu\nu}$, more precisely the Hubble parameter and its cosmic time derivatives as defined from the metric $g_{\mu\nu}$, near the singularities is consistent with that of the geodesic behaviour dictated by the same metric $g_{\mu\nu}$. However, the events corresponding to the singularities with respect to $g_{\mu\nu}$ are usually well behaved as observed by the connection, and therefore the auxiliary metric, and so do the geodesic equations defined from the physical connection. In addition, we show that bound structures close to a big rip or little rip will be destroyed before the advent of the singularity and will remain bound close to a sudden, big freeze or type IV singularity. This result is independent of the choice of the physical or auxiliary metric.

We will further complete our analyses by getting some observational constraints on the model through the use of a cosmographic approach \cite{Capozziello:2008qc,Capozziello:2009ka,BouhmadiLopez:2010pp,Capozziello:2011tj}. This analysis will show that the EiBI model when filled with the matter content mentioned on the previous paragraph on top of the dark and baryonic matter is compatible with the current acceleration of the Universe. The cosmographic approach relies on putting constraints on some parameters which quantify the time derivatives of the scale factor and which are called the cosmographic parameters \cite{Capozziello:2008qc,Capozziello:2009ka,BouhmadiLopez:2010pp,Capozziello:2011tj}. These parameters depend exclusively on the space-time geometry, in this case on the geometry of a  homogeneous and isotropic space-time, and not on the gravitational action or the equations of motion that describe the model (see Ref.~\cite{Capozziello:2009ka} for a nice review on the subject). Hence, this approach is quite useful because given a set of constraints on the cosmographic parameters \cite{Capozziello:2008qc,Capozziello:2011tj}, it can be applied to a large amount of models in particular to those with relatively messy Friedmann equations like the one we need to deal with \cite{Avelino:2012ue}. The drawback of this approach is that with the current observational data the errors can be quite large \cite{Capozziello:2008qc,Capozziello:2011tj,Aviles:2012ay,Gruber:2013wua,Vitagliano:2009et,Xia:2011iv,Lazkoz:2013ija}. Nevertheless, we think it is a fear enough approach for the analysis we want to carry out. Essentially, We will show that among the various birth events or past singularities predicted by the theory, the cosmographic analyses pick up the physical region which determines the occurrence of a type IV singularity (or a loitering effect) in the past, which is the most unharmful of all the types of dark energy singularities. While among the various possible future singularities or doomsdays predicted, the use of observational constraints on higher order cosmographic parameters is necessary to predict which future singularity is more probable.

The paper is outlined as follows. In section II, we shortly review the idea of the EiBI theory and present a thorough analysis on the avoidance of various singularities in this theory, through deriving the asymptotic behaviours of the Hubble parameter and its cosmic time derivatives near the singularities for both metrics (physical and auxiliary). In section III, we analyse the effects of the cosmological expansion on local bound systems in the EiBI scenario by analysing the geodesics of test particles for both metrics close to a massive body. In section IV, we use a cosmographic approach to constrain the model, and calculate the cosmic time elapsed since now to the possible, past or future, singularities. The conclusions and discussions are presented in section V. 

\section{The EiBI model and dark energy related singularities}\label{sectII}

We start reviewing the EiBI model whose gravitational action in terms of the metric $g_{\mu\nu}$ and the connection $\Gamma^{\alpha}_{\mu\nu}$
reads \cite{Banados:2010ix} 
\begin{eqnarray}
\mathcal{S}_{\textrm{EiBI}}(g,\Gamma,\Psi)&=&\frac{2}{\kappa}\int d^4x\left[\sqrt{|g_{\mu\nu}+\kappa R_{\mu\nu}(\Gamma)|}-\lambda\sqrt{|g|}\right] \nonumber\\ &\,&+\,\mathcal{S}_\textrm{m} (g,\Psi).
\label{action}
\end{eqnarray} 
The theory is formulated within the Palatini approach and therefore the Ricci tensor is purely constructed from the connection $\Gamma$. In addition, $R_{\mu\nu}(\Gamma)$ in the action \eqref{action} is chosen to be the symmetric part of the Ricci tensor and the connection is also assumed to be torsionless. Within the Palatini formalism we are assuming here, the connection $\Gamma^{\alpha}_{\mu\nu}$ and the metric $g_{\mu\nu}$ are treated as independent variables. The parameter $\kappa$ is a constant with inverse dimensions to that of a cosmological constant (in this paper, we will work with Planck units $8\pi{G}=1$ and set the speed of light to $c=1$), $\lambda$ is a dimensionless constant and $\mathcal{S}_\textrm{m} (g,\Psi)$ stands for the matter Lagrangian in which matter is assumed to be coupled covariantly to the metric $g$ only. Therefore, the energy momentum tensor derived from Eq.~\eqref{action} is conserved like in GR \cite{Banados:2010ix}. One can also note that the action \eqref{action} will recover the Einstein-Hilbert action as $|\kappa R|$ gets very small with an effective cosmological constant $\Lambda=(\lambda-1)/\kappa$ \cite{Banados:2010ix}. From now on we will assume a vanishing effective cosmological constant, i.e., $\lambda=1$. In addition, we will restrict our analysis to a positive $\kappa$, in order to avoid the imaginary effective sound speed instabilities usually present in the EiBI theory with negative $\kappa$ \cite{Avelino:2012ge}. 

For a FLRW universe filled with a perfect fluid with energy density $\rho$ and pressure $p$, the Friedmann equation reads \cite{Avelino:2012ue}
\begin{eqnarray}
\bar{H}^2&=&\frac{8}{3}\Big[\bar\rho+3\bar p-2+2\sqrt{(1+\bar\rho)(1-\bar p)^3}\Big]\nonumber\\
&\times&\frac{(1+\bar\rho)(1-\bar p)^2}{[(1-\bar p)(4+\bar\rho-3\bar p)+3\frac{d\bar p}{d\bar\rho}(1+\bar\rho)(\bar\rho+\bar p)]^2},\nonumber\\
\label{field equation}
\end{eqnarray}
where $\bar H\equiv\sqrt{\kappa}H$, $H$ is the Hubble parameter as defined from the physical metric, $\bar\rho=\kappa\rho$, $\bar p=\kappa p$, and $d\bar p/d\bar\rho\equiv c_s^2$ denotes the derivative of the pressure with respect to the energy density. For simplicity, we will also use the following dimensionless cosmic time: $\bar t\equiv t/\sqrt{\kappa}$ where $t$ corresponds to the cosmic time as defined from the physical metric $g_{\mu\nu}$. When the curvature gets very small, i.e., $|\kappa R|\ll|g|$, the Friedmann equation \eqref{field equation} becomes
\begin{equation}
{\bar H}^2\approx\frac{\bar\rho}{3}-\frac{3w^2+2w-15}{8}(\bar\rho)^2+\textrm{higher order of }\bar\rho,
\end{equation}
where a constant equation of state $\bar p=w\bar\rho$ is considered~\footnote{The leading order in the expansion of the scalar curvature with respect to $\bar\rho$ satisfies $\kappa R\propto\bar\rho$ at the low energy density limit, thus we can expand with respect to the energy density when the low curvature assumption is considered.}. Recall that the EiBI theory recovers GR when $|\kappa R|$ is very small as shown in \cite{Banados:2010ix}. On the other hand, the conservation equation, as mentioned previously, takes the standard form
\begin{equation}
\frac{d\bar\rho}{d\bar t}+3 \bar H(\bar\rho+\bar p)=0.
\label{conservation equation}
\end{equation}

It can be easily verified that the Big Bang singularity can be avoided in this theory for a radiation dominated universe \cite{Banados:2010ix}; i.e. $\bar p=\bar \rho/3$, and in general a universe filled with a perfect fluid with a constant and positive equation of state $w$; i.e., fulfilling the null energy conditions \cite{largescale}, bounces in the past for $\kappa<0$ or has a loitering behaviour in the infinite past for $\kappa>0$ \cite{Scargill:2012kg}.

Aside, we can define an auxiliary metric $q_{\mu\nu}$ which is compatible with the connection $\Gamma$ \cite{Banados:2010ix}:
\begin{equation}
q_{\mu\nu}dx^\mu dx^\nu=-U(t)dt^2+a^2(t)V(t)(dx^2+dy^2+dz^2),
\end{equation}
where 
\begin{eqnarray}
U&=&\sqrt{\frac{(1-\bar p)^3}{1+\bar\rho}},\\
V&=&\sqrt{(1+\bar\rho)(1-\bar p)},\label{V}
\end{eqnarray}
and $a$ is the scale factor of the physical metric $g_{\mu\nu}$. From the auxiliary metric $q_{\mu\nu}$ we can define as well an auxiliary Hubble parameter $H_q$ whose rescaled dimensionless value can be expressed as $\bar{H}_q\equiv\sqrt{\kappa}H_q$ and reads
\begin{eqnarray}
\bar{H}_q=\sqrt{\kappa}\frac{1}{\tilde{a}}\frac{d\tilde{a}}{d\tilde t}=\frac{1}{\sqrt{U}}\frac{d}{d\bar{t}}\ln(a\sqrt{V}),
\label{defHq}
\end{eqnarray}
where $\tilde a\equiv\sqrt{V}a$ and $d\tilde t\equiv\sqrt{U}dt$. Besides, we find that $\bar H_q$ satisfies
\begin{eqnarray}
\kappa q^{\mu\nu}R_{\mu\nu}(\Gamma)&=&12{\bar{H}_q}^2+6\sqrt{\kappa}\frac{d\bar{H}_q}{d\tilde t}\nonumber\\
&=&4-\frac{1}{U}-\frac{3}{V},
\label{aux}
\end{eqnarray}
where
\begin{equation}
{\bar H_q}^2=\frac{1}{3}+\frac{\bar\rho+3\bar p-2}{6\sqrt{(1+\bar\rho)(1-\bar p)^3}}.
\label{HHq}
\end{equation}
Notice that ${\bar H_q}^2$ does not depend on $c_s^2$, unlike ${\bar H}^2$ in Eq.~\eqref{field equation}. One can see that this auxiliary Hubble parameter also recovers the Hubble parameter in standard GR as the curvature gets small:
\begin{equation}
{\bar H_q}^2\approx\frac{\bar\rho}{3}+\frac{3w^2+6w-5}{24}(\bar\rho)^2+\textrm{higher order of }\bar\rho,
\end{equation}
where $w$ is also a constant equation of state parameter. This auxiliary metric which is compatible with the physical connection cannot avoid the Big Bang singularity in the past because both $H_q$ and $dH_q/d\tilde t$ diverge at a vanishing $\tilde a$ and at a finite past $\tilde t$, and so does the Ricci scalar defined in Eq.~(\ref{aux}).

We will next analyse the possible avoidance of dark energy singularities in the EiBI setup. Those singularities, as we will next review, are characterised by possible divergence of the Hubble parameter and its cosmic time derivatives at some finite cosmic time. This translates into possible divergences of the scalar curvature and its cosmic time derivatives. The EiBI model we are considering is formulated within the Palatini formalism and therefore there are two ways of defining the Ricci curvature: (i) $R_{\mu\nu}(\Gamma)$ as presented in the action (\ref{action})  and (ii) $R_{\mu\nu}(g)$ constructed from the metric $g_{\mu\nu}$. There are in addition four ways of defining the scalar curvature: $g^{\mu\nu}R_{\mu\nu}(\Gamma)$, $g^{\mu\nu}R_{\mu\nu}(g)$, $q^{\mu\nu}R_{\mu\nu}(\Gamma)$ and $q^{\mu\nu}R_{\mu\nu}(g)$. Therefore whenever one refers to singularity avoidance, one must specify the specific curvature one is referring to. For the dark energy singularities the important issue is the behaviour of the Hubble parameter and its cosmic time derivatives and in this case we have two possible quantities for the Hubble parameter: $H$ related to the physical metric and $H_q$ related to the physical connection as defined in Eq.~(\ref{defHq}). 

In general, the Universe is filled with radiation, dark and baryonic matter, and dark energy:
\begin{eqnarray}
\bar\rho=\bar\rho_r+\bar\rho_m+\bar\rho_{de},\nonumber\\
\bar p=\frac{1}{3}\bar\rho_r+\bar p_{de}(\bar\rho_{de}),\
\label{content}
\end{eqnarray}
where $\bar\rho_r=\kappa\rho_r$, $\bar\rho_{m}=\kappa\rho_{m}$, $\bar\rho_{de}=\kappa\rho_{de}$, and $\bar p_{de}=\kappa p_{de}$ are the energy density of radiation, matter, dark energy and the pressure of dark energy, respectively. Note that $p_{de}(\bar\rho_{de})$ means that the equation of state of dark energy is purely a function of the dark energy density. For the sake of completeness, we will assume a universe filled with a matter contents as shown in Eq.~\eqref{content} to go through the analysis in this paper. Note that even though dark matter and radiation are unimportant for the analysis of future singularities, they are not for the analysis of past singularities.

Before starting our analysis, we will review the definition of these dark energy related singularities:

\begin{itemize}

\item The Big Rip singularity happens at a finite cosmic time with an infinite scale factor where the Hubble parameter and its cosmic time derivative diverge \cite{Starobinsky:1999yw,Caldwell:2003vq,Caldwell:1999ew,Carroll:2003st,Chimento:2003qy,Dabrowski:2003jm,GonzalezDiaz:2003rf,GonzalezDiaz:2004vq}

\item The Sudden singularity takes place at a finite cosmic time with a finite scale factor, where the Hubble parameter remains finite but its cosmic time derivative diverges \cite{Barrow:2004xh,Gorini:2003wa,Nojiri:2005sx}.

\item The Big Freeze singularity happens at a finite cosmic time with a finite scale factor where the Hubble parameter and its cosmic time derivative diverge \cite{BouhmadiLopez:2007qb,BouhmadiLopez:2006fu,Nojiri:2005sx,Nojiri:2004pf,Nojiri:2005sr}.

\item Finally Type \rom{4} singularity occurs at a finite cosmic time with a finite scale factor where the Hubble parameter and its cosmic time derivative remain finite, but higher cosmic time derivatives of the Hubble parameter still diverge \cite{BouhmadiLopez:2006fu,Nojiri:2005sx,Nojiri:2004pf,Nojiri:2005sr,Nojiri:2008fk,Bamba:2008ut}.

\end{itemize}

To analyse the Big Freeze, Sudden, and Type IV singularities, we regard the phantom Generalized Chaplygin Gas (pGCG) as the dark energy component in this model \cite{BouhmadiLopez:2007qb,BouhmadiLopez:2004me}. Its equation of state takes the form:
\begin{equation}
\bar p_{de}=-\frac{A}{(\bar\rho_{de})^{\alpha}},
\label{pgcg EOM}
\end{equation}
where $\alpha$ and $A>0$ are two dimensionless constants. In GR, this kind of phantom energy will drive a past sudden singularity for $\alpha>0$, a future big freeze singularity for $\alpha<-1$, and a past type IV singularity for $-1<\alpha<0$ except for some quantized values of $\alpha$ in which the Hubble rate and its higher order derivatives are all regular in the finite past \cite{BouhmadiLopez:2007qb}. Note that the last case is different from the results shown in Ref.~\cite{BouhmadiLopez:2007qb} because in that reference the authors assumed a universe filled only with a pGCG instead of the matter content given in Eq.~\eqref{content} to which we will stick in this paper. Actually, the addition of radiation and matter contributions does not make any comparable difference to the cases in which past Sudden and future Big Freeze occur in GR, i.e., $\alpha>0$ and $\alpha<-1$, respectively. However, the conclusion is different when $-1<\alpha<0$. {\color{black}See Ref.~\cite{Bouhmadi-Lopez:2014jfa} for more details on this issue.}

After integrating the conservation equation (\ref{conservation equation}) and assuming $\alpha>-1$, one can derive the energy density of this kind of pGCG which drives the finite past Sudden or Type IV singularity in GR \cite{BouhmadiLopez:2007qb}:
\begin{equation}
\bar\rho_{de}=A^{\frac{1}{1+\alpha}}\left[1-\left(\frac{a}{a_\textrm{min}}\right)^{-3(1+\alpha)}\right]^{\frac{1}{1+\alpha}},
\label{past density}
\end{equation}
where $a_\textrm{min}$ is the scale factor corresponding to the singularity. 

For later convenience, we also rewrite the energy density in terms of the scale factor as
\begin{equation}
\bar\rho_{de}=\bar\rho_{de0}\left[\frac{1-\left(\frac{a_\textrm{min}}{a}\right)^{3(1+\alpha)}}{1-{a_\textrm{min}}^{3(1+\alpha)}}\right]^{\frac{1}{1+\alpha}}.
\label{rpp}
\end{equation}
Note here that we have set the scale factor at present, $a_0$, as $a_0=1$ and we will use this convention in the rest of this paper. A subscript $0$ stands for quantities evaluated today. On the other hand, if $\alpha<-1$ and $A>0$, the energy density of this pGCG which drives the finite future Big Freeze singularity in GR reads \cite{BouhmadiLopez:2007qb}:
\begin{equation}
\bar\rho_{de}=A^{\frac{1}{1+\alpha}}\left[1-\left(\frac{a}{a_\textrm{max}}\right)^{-3(1+\alpha)}\right]^{\frac{1}{1+\alpha}},
\label{future density}
\end{equation}
where $a_\textrm{max}$ is the scale factor corresponding to the future singularity.

We also rewrite the energy density in terms of the scale factor as follows for the sake of later convenience:
\begin{equation}
\bar\rho_{de}=\bar\rho_{de0}\left[\frac{1-\left(\frac{a_\textrm{max}}{a}\right)^{3(1+\alpha)}}{1-{a_\textrm{max}}^{3(1+\alpha)}}\right]^{\frac{1}{1+\alpha}}.
\label{rfs}
\end{equation}

Additionally, there are some special case in which the phantom character shares the same equation of state \eqref{pgcg EOM} while does not imply $A>0$, as shown in Refs.~\cite{BouhmadiLopez:2007qb,BouhmadiLopez:2004me}. This special pGCG will drive a finite future big freeze singularity in GR and its energy density and pressure are
\begin{eqnarray}
\bar\rho_{de}&=&|A|^{\frac{1}{1+\alpha}}\left[\left(\frac{a}{a_\textrm{max}}\right)^{-3(1+\alpha)}-1\right]^{\frac{1}{1+\alpha}},\nonumber\\
\bar p_{de}&=&-\frac{A}{(\bar\rho_{de})^{\alpha}}\nonumber\\
&=&|A|^{\frac{1}{1+\alpha}}\left[\left(\frac{a}{a_\textrm{max}}-1\right)^{-3(1+\alpha)}-1\right]^{\frac{1}{1+\alpha}-1},\nonumber\\
\label{specialpgcg}
\end{eqnarray}
where $A<0$ and $1+\alpha=1/(2m)$ with $m$ being a negative integer \cite{BouhmadiLopez:2007qb}. We will also discuss this special case within the EiBI scenario in the upcoming subsection.

\subsection{The EiBI scenario and the Big Rip}
\subsubsection{The physical metric $g_{\mu\nu}$}

We showed recently that despite the Big Bang avoidance in the EiBI setup, the Big Rip singularity  \cite{Starobinsky:1999yw,Caldwell:2003vq} is unavoidable in the EiBI phantom model \cite{Bouhmadi-Lopez:2013lha}. Indeed, we have shown analytically and numerically that in the EiBI theory, a universe filled with matter and phantom energy with a constant equation of state $w<-1$ will still hit a big rip singularity; i.e. the Hubble parameter $\bar H$ and $d{\bar H}/d{\bar t}$ blow up in a finite future cosmic time and at an infinite scale factor. Essentially, the square of the dimensionless Hubble parameter $\bar H$ and its cosmic time derivative near the singularity are almost linear functions of the energy density:
\begin{eqnarray}
{\bar H}^2&\approx&\frac{4\sqrt{|w|^3}}{3(3w+1)^2}\bar\rho\rightarrow\infty,\nonumber\\
\frac{d\bar H}{d\bar t}&\approx&\frac{2\sqrt{|w|^3}}{(3w+1)^2}|1+w|\bar\rho\rightarrow\infty.
\label{HH in Big Rip}
\end{eqnarray}
Therefore, at very large scale factor and energy density (which grows as $\bar\rho \propto a^{-3(1+w)}$ for $w<-1$ and constant), $\bar H$ and $d{\bar H}/d{\bar t}$ get equally large. This happens at a finite future cosmic time \cite{Bouhmadi-Lopez:2013lha}.

\subsubsection{The auxiliary metric $q_{\mu\nu}$}

As for the quantities defined by the auxiliary metric, it can be shown that 
\begin{eqnarray}
{\bar H_q}^2&\approx&\frac{1}{3}+\frac{1+3w}{6\sqrt{|w|^3}\bar\rho}\rightarrow\frac{1}{3},\nonumber\\
\sqrt{\kappa}\frac{d{\bar H}_q}{d\tilde t}&\approx&\frac{|1+w|}{2\sqrt{|w|^3}\bar\rho}\rightarrow0,
\end{eqnarray}
and second and higher order derivatives of $\bar H_q$ with respect to $\tilde t$ vanish when $\bar\rho\rightarrow\infty$ because their leading order in the expansion on $\bar\rho$ is inversely proportional to $\bar\rho$. Furthermore, we also find that the energy density blows up and
\begin{equation}
\tilde a\propto e^{H_q\tilde t}
\end{equation}
when $\tilde t\rightarrow\infty$, which corresponds to a finite $t$. Therefore, there is no singularity when the auxiliary metric is considered to be on the form of a FLRW metric in the EiBI theory. Indeed, the Universe approaches a de Sitter state as described by the auxiliary metric in this case. Note that according to Eq.~(\ref{aux}), $q^{\mu\nu}R_{\mu\nu}(\Gamma)\approx4/\kappa$ as $\bar\rho\rightarrow\infty$, which is in concordance with our previous results \cite{Bouhmadi-Lopez:2013lha}.

\subsection{The EiBI scenario and the Sudden singularity}
\subsubsection{The physical metric $g_{\mu\nu}$}

We seek now the possibility of smoothing the Sudden singularity that can appear in GR. We consider a pGCG fulfilling the equation of state \eqref{pgcg EOM} with $\alpha>0$ \cite{BouhmadiLopez:2007qb}. Note that in GR a universe filled with this fluid hits a past sudden singularity. The presence of matter or radiation cannot remove the occurrence of this cosmic birth on the past of the Universe.  After integrating the conservation equation (\ref{conservation equation}), one can derive the energy density of this kind of pGCG \cite{BouhmadiLopez:2007qb} which is shown in Eqs.~\eqref{past density} and \eqref{rpp}.

As the Universe is filled with radiation, matter and pGCG with $\alpha>0$, the asymptotic behaviour of ${\bar H}^2$ and its cosmic time derivatives near the singularity ($a\rightarrow a_\textrm{min}$, $\bar\rho_{de}\rightarrow0$, and $\bar\rho\rightarrow\bar\rho_r+\bar\rho_m\rightarrow\bar\rho_\textrm{ini}$ where $\bar\rho_{\textrm{ini}}$ is the initial dimensionless energy density at $a=a_{\textrm{min}}$) are the following:
\begin{eqnarray}
{\bar H}^2&\approx&\frac{16(\bar\rho_{de})^{2+\frac{\alpha}{2}}}{27\sqrt{A(1+\bar\rho_\textrm{ini})}\alpha^2}+\textrm{higher order of }\bar\rho_\textrm{de},\nonumber\\
\frac{d\bar H}{d\bar t}&\approx&\sqrt{\frac{A}{1+\bar\rho_\textrm{ini}}}\frac{4(\alpha+4)(\bar\rho_{de})^{1-\frac{\alpha}{2}}}{9\alpha^2}\nonumber\\
&+&\textrm{higher order of }\bar\rho_\textrm{de},
\label{HH sudden neq2}
\end{eqnarray}
for $\alpha\neq2$; more precisely, we find that
\begin{widetext}
\begin{equation}
{\bar H}^2\approx\left\{
\begin{array}{ll}
\frac{16(\bar\rho_{de})^{2+\frac{\alpha}{2}}}{27\sqrt{A(1+\bar\rho_\textrm{ini})}\alpha^2}\Bigg(1-\frac{3}{2}\frac{1}{
\sqrt{A(1+\bar\rho_\textrm{ini})}}(\bar\rho_\textrm{de})^{\frac{1}{2}\alpha}+\textrm{higher order of }\bar\rho_\textrm{de}\Bigg) & 0<\alpha< 2 \\
\frac{16(\bar\rho_{de})^{2+\frac{\alpha}{2}}}{27\sqrt{A(1+\bar\rho_\textrm{ini})}\alpha^2}\Bigg(1+\frac{1}{(1+\bar\rho_\textrm{ini})}\frac{4-\alpha}{2\alpha}(\bar\rho_\textrm{de})+\textrm{higher order of }\bar\rho_\textrm{de}\Bigg) & \alpha>2
\end{array},\right.
\end{equation}
\end{widetext}
and
\begin{eqnarray}
{\bar H}^2&\approx& b_3(\bar\rho_{de})^3+b_4(\bar\rho_{de})^4+b_5(\bar\rho_{de})^5+O^6(\bar\rho_{de}),\nonumber\\
\frac{d\bar H}{d\bar t}&\approx&\frac{9}{2}b_3A+6b_4A\bar\rho_{de}+O^2(\bar\rho_{de}),\nonumber\\
\frac{d^2\bar H}{d{\bar t}^2}&\approx&6b_4A^2\sqrt{\frac{b_3}{\bar\rho_{de}}}\rightarrow\infty,
\end{eqnarray}
for $\alpha=2$, where
\begin{eqnarray}
b_3&=&\frac{4}{27\sqrt{1+\bar\rho_\textrm{ini}}}A^{-\frac{1}{2}},\nonumber\\
b_4&=&\frac{2}{27\sqrt{(1+\bar\rho_\textrm{ini})^3}}A^{-\frac{1}{2}}-\frac{2}{9(1+\bar\rho_\textrm{ini})}A^{-1},\nonumber\\
b_5&=&-\frac{1}{18\sqrt{(1+\bar\rho_\textrm{ini})^5}}A^{-\frac{1}{2}}+\frac{14(3-\bar\rho_r)}{81\sqrt{1+\bar\rho_\textrm{ini}}}A^{-\frac{3}{2}}.\nonumber\\
\end{eqnarray}
One can see that the first cosmic time derivative of the Hubble rate blows up if $\alpha>2$ and the second order derivative of the Hubble rate blows up if $\alpha=2$ and $b_4\neq 0$ because $\bar\rho_{de}$ vanishes at $a=a_\textrm{min}$. If $\alpha=2$ and $b_4=0$, the second order cosmic time derivative is finite, but the third order derivative diverges as $\bar\rho_{de}\rightarrow 0$.

The scale factor dependence on the cosmic time since the Universe expands from $a_\textrm{min}$ to a given size (at a given $\bar t$) can be approximated as follows:
\begin{equation}
\frac{a}{a_\textrm{min}}\approx1+\left\{\frac{3\alpha D}{4(1+\alpha)}\left[3A(\alpha+1)\right]^{\frac{\alpha+4}{4(\alpha+1)}}(\bar t-\bar t_\textrm{min})\right\}^{\frac{4(\alpha+1)}{3\alpha}},
\label{cosmic time neq 2}
\end{equation}
where $D=4/[3\alpha\sqrt{3(A(1+\bar\rho_\textrm{ini}))^{1/2}}]$ for $\alpha\neq2$; and
\begin{equation}
\frac{a}{a_\textrm{min}}\approx1+\left[\frac{3}{2}\sqrt{b_3A}(\bar t-\bar t_\textrm{min})\right]^2,
\end{equation}
for $\alpha=2$. One find that the Universe starts expanding from a finite past in these cases. Therefore the Universe hits a sudden singularity for $\alpha>2$ and a type IV singularity for $\alpha=2$.

Furthermore, if $0<\alpha<2$ and $\alpha=4/(3n+2)$ in which $n$ is a natural number, the $(n+2)$-th derivative of $\bar H$ will diverge even though the $1$,...,$(n+1)$-th derivatives are all regular. The reason is the following: the $(n+1)$-th derivative of $\bar H$ behaves as
\begin{equation}
\frac{d^{n+1}}{d{\bar t}^{n+1}}\bar H\propto C_{n+1}+(\bar\rho_{de})^{\frac{\alpha}{2}}+\textrm{higher order of }\bar\rho_{de},
\end{equation}
with $C_{n+1}$ being a finite non-vanishing constant. Then the next order becomes
\begin{equation}
\frac{d^{n+2}}{d{\bar t}^{n+2}}\bar H\propto (\bar\rho_{de})^{-\frac{\alpha}{4}}+\textrm{higher order of }\bar\rho_{de},
\end{equation}
which diverges because $\alpha>0$ and implies a type IV singularity in the finite past. 

If, however, $0<\alpha<2$ and $\alpha\neq4/(3n+2)$, we find that as long as $\alpha$ satisfies $4/(3p+2)<\alpha<4/(3p-1)$ with $p$ being a positive integer, the $(p+1)$-th derivative of $\bar H$ blows up while the $1$,...,$p$-th derivatives are all finite. This indicates a type IV singularity again.

Hence, the past Sudden singularity originally driven by a pGCG in GR will induce the following behaviours in the EiBI scenario:

\begin{itemize}
\item If $\alpha>2$, the Universe expands from a finite past sudden singularity.
\item If $0<\alpha\le2$, the Universe expands from a finite past type IV singularity.
\end{itemize}

Actually, the Sudden singularities can also occur in the future if the Universe is filled with a GCG which fulfils the strong, null, and weak energy conditions within the GR setup \cite{BouhmadiLopez:2007qb}. However, the Universe will not get into a late-time accelerating expansion phase that is observationally corroborated, so that the theory will only be worth analysing from a mathematical point of view. {\color{black}See Ref.~\cite{Bouhmadi-Lopez:2014jfa} for more details on this issue.}

\subsubsection{The auxiliary metric $q_{\mu\nu}$}

On the other hand, it can be shown for $\alpha>0$ and $A>0$ that
\begin{eqnarray}
{\bar H_q}^2&\approx&\frac{1}{3}-\frac{(\bar\rho_{de})^{\frac{\alpha}{2}}}{2\sqrt{A(1+\bar\rho_\textrm{ini})}}\rightarrow\frac{1}{3},\nonumber\\
\sqrt{\kappa}\frac{d{\bar H}_q}{d\tilde t}&\approx&
\frac{(\bar\rho_{de})^{\frac{\alpha}{2}}}{2\sqrt{A(1+\bar\rho_\textrm{ini})}}\rightarrow0,
\end{eqnarray}
and second and higher order derivatives of $\bar H_q$ with respect to $\tilde t$ vanish when $a\rightarrow a_\textrm{min}$, as well as $\bar\rho_{de}\rightarrow0$, because their leading order in the expansion on $\bar\rho_{de}$ is proportional to $(\bar\rho_{de})^{\alpha/2}$. Notice that in this case the auxiliary Hubble rate $\bar H_q$ is negative when $\bar\rho_{de}\rightarrow0$ in the past because $\tilde a=\sqrt{V}a\rightarrow+\infty$ because $V\rightarrow\infty$ (see Eqs.~\eqref{V} and \eqref{pgcg EOM} for $a\rightarrow a_\textrm{min}$) and this happens at an infinite past $\tilde t$. Furthermore, we also find that
\begin{equation}
\tilde a\propto e^{H_q\tilde t}\textrm{ when }\tilde t\rightarrow-\infty.
\end{equation}
Indeed, the Universe approaches a contracting de Sitter state as described by the auxiliary metric in this case. Therefore, there is no singularity of the auxiliary metric when the auxiliary metric is considered to be into a FLRW form within the EiBI theory.
Note that $q^{\mu\nu}R_{\mu\nu}(\Gamma)\approx4/\kappa$ when $a\rightarrow a_\textrm{min}$ and $\tilde a\rightarrow\infty$ in this case.

\subsection{The EiBI scenario and the Big Freeze}
\subsubsection{The physical metric $g_{\mu\nu}$}

We seek now the possibility of smoothing the Big Freeze singularity that can appear in GR. We consider a pGCG fulfilling the equation of state \eqref{pgcg EOM} with $\alpha<-1$ \cite{BouhmadiLopez:2007qb}. Note that in GR a universe filled with this fluid hits a future big freeze singularity. The presence of matter or radiation cannot remove the occurrence of this cosmic doomsday on the future of the Universe.  After integrating the conservation equation (\ref{conservation equation}), one can derive the energy density of this kind of pGCG \cite{BouhmadiLopez:2007qb} which is shown in Eqs.~\eqref{future density} and \eqref{rfs}.

The asymptotic behaviour of ${\bar H}^2$ and the cosmic time derivatives of the Hubble parameter as $a\rightarrow a_\textrm{max}$, $\bar\rho\approx\bar\rho_{de}\rightarrow\infty$ within the EiBI setup reads
\begin{eqnarray}
{\bar H}^2&\approx&\frac{16(\bar\rho)^{\frac{3+\alpha}{2}}}{27(1-\alpha)^2\sqrt{A}}+\textrm{higher order of }(\bar\rho)^{-1},\nonumber\\
\frac{d\bar H}{d\bar t}&\approx&\frac{4(\alpha+3)\sqrt{A}(\bar\rho)^{\frac{1-\alpha}{2}}}{9(1-\alpha)^2}+\textrm{higher order of }(\bar\rho)^{-1},\nonumber\\
\label{big freeze beq -3}
\end{eqnarray}
for $\alpha\neq-3$; more precisely, we find that
\begin{widetext}
\begin{equation}
{\bar H}^2\approx\left\{
\begin{array}{ll}
\frac{16(\bar\rho)^{\frac{3+\alpha}{2}}}{27(1-\alpha)^2\sqrt{A}}\Bigg(1-\frac{2}{3}\left(\frac{1+3\alpha}{1-\alpha
}\right)\frac{1}{A}(\bar\rho)^{1+\alpha}+\textrm{higher order of }(\bar\rho)^{-1}\Bigg) & -2<\alpha< -1 \\
\frac{16(\bar\rho)^{\frac{3+\alpha}{2}}}{27(1-\alpha)^2\sqrt{A}}\Bigg(1+\frac{3+\alpha}{2(1-\alpha)}(1+\bar\rho_m+\bar\rho_r)(\bar\rho)^{-1}+\textrm{higher order of }(\bar\rho)^{-1}\Bigg) & \alpha<-2 \\
\frac{16(\bar\rho)^{\frac{1}{2}}}{243\sqrt{A}}\Bigg[1+\Bigg(\frac{1}{6}(1+\bar\rho_m+\bar\rho_r)+\frac{10}{9}\frac{1}{A}\Bigg)\
(\bar\rho)^{-1}+\textrm{higher order of }(\bar\rho)^{-1}\Bigg] & \alpha=-2
\end{array},\right.
\end{equation}
\newpage
\end{widetext}
and
\begin{eqnarray}
{\bar H}^2&\approx& c_0+c_2(\bar\rho)^{-2}+c_3(\bar\rho)^{-3}+O^{-4}(\bar\rho),\nonumber\\
\frac{d\bar H}{d\bar t}&\approx&-3 c_2A-\frac{9}{2}c_3 A(\bar\rho)^{-1}+O^{-2}(\bar\rho),\nonumber\\
\frac{d^2\bar H}{d{\bar t}^2}&\approx&\frac{27}{2}c_3A^2\bar H\bar\rho\rightarrow\infty.
\end{eqnarray}
for $\alpha=-3$, where
\begin{eqnarray}
c_0&=&\frac{1}{27}A^{-\frac{1}{2}},\nonumber\\
c_2&=&-\frac{(1+\bar\rho_m+\bar\rho_r)^2}{144}A^{-\frac{1}{2}}-\frac{1}{18}A^{-1}+\frac{4}{81}A^{-\frac{3}{2}},\nonumber\\
c_3&=&\frac{7(1+\bar\rho_m+\bar\rho_r)^3}{864}A^{-\frac{1}{2}}+\frac{(1+\bar\rho_m+\bar\rho_r)}{36}A^{-1}\nonumber\\
&+&\frac{(17+2\bar\rho_m-3\bar\rho_r)}{162}A^{-\frac{3}{2}},
\end{eqnarray}
in which $\bar\rho_{m}$ and $\bar\rho_{r}$ denote the dimensionless energy density of matter and radiation at $a=a_\textrm{max}$.
It can be shown that $c_0$ and $c_3$ are always positive for any physical value of $A$ (note that $\bar\rho_r\ll1$ at $a=a_\textrm{max}$). Nevertheless, $c_2$ can vanish but the first derivative of $\bar H$ with respect to the cosmic time is still finite, more precisely, it vanishes when $\bar\rho$ blows up.

It can also be shown that the scale factor dependence on the cosmic time since the Universe has a given size (at a given $\bar t$) till it reaches $a_\textrm{max}$ is the following:
\begin{eqnarray}
\bar t_\textrm{max}-\bar t&\,\approx& \,\frac{1}{\tilde D(3A|\alpha+1|)^{\frac{\alpha+3}{4(\alpha+1)}}}\left[\frac{4(\alpha+1)}{1+3\alpha}\right]\times \nonumber\\
 &\,&\,\,\,\,\,\,\,\,\,\,\,\,\left(1-\frac{a}{a_\textrm{max}}\right)^{\frac{1+3\alpha}{4(1+\alpha)}},
\label{alpha not -3 t}
\end{eqnarray}
where $\tilde D=4/[3(1-\alpha)\sqrt{3(A)^{1/2}}]$ for $\alpha\neq-3$; and
\begin{equation}
\frac{a}{a_{\textrm{max}}}\approx 1-\sqrt{c_0}(\bar t_\textrm{max}-\bar t),
\label{alpha -3 t}
\end{equation}
for $\alpha=-3$.

One can find from Eqs.~(\ref{alpha not -3 t}) and (\ref{alpha -3 t}) that the cosmic time till the scale factor approaches $a_\textrm{max}$ is finite for $\alpha<-1$. Therefore, a pGCG with $\alpha<-1$, that leads to a big freeze in GR, fuels the following behaviour in the EiBI setup:

\begin{itemize}
\item If $\alpha<-3$, the Universe will  end up into a finite future sudden singularity.
\item If $-3<\alpha<-1$, the Universe will end up into a finite future big freeze singularity.
\item If $\alpha=-3$, the Universe will end up into a finite future type IV singularity. 
\end{itemize}
In summary, as with respect to GR ($\alpha<-1$) the Big Freeze singularity is smoothed in general except for $-3<\alpha<-1$ which maintains its GR character.

Additionally, there is also a finite future big freeze singularity in GR, which is driven by a very special pGCG whose energy density and pressure are shown in Eqs.~\eqref{specialpgcg},
where $A<0$ and $1+\alpha=1/(2m)$ with $m$ being a negative integer \cite{BouhmadiLopez:2007qb}. The asymptotic behaviour of ${\bar H}^2$ and $d\bar H/d\bar t$ on this case are also given by Eqs.~(\ref{big freeze beq -3}). One can easily see that $-3<\alpha<-1$, thus the Big Freeze singularity cannot be avoided in this case.

Actually, the Big Freeze singularity can also occur in the finite past if the Universe is filled with a GCG which fulfils the strong, null, and weak energy conditions \cite{BouhmadiLopez:2007qb}. However, the Universe will not get into an accelerating expansion phase at the present time as implied by astrophysical and cosmological observations, so that the theory will only be worth to be analysed from a mathematical point of view. {\color{black}See Ref.~\cite{Bouhmadi-Lopez:2014jfa} for more details on this issue.}

\subsubsection{The auxiliary metric $q_{\mu\nu}$}

On the other hand, it can also be shown that for $\alpha<-1$ and $A>0$ 
\begin{eqnarray}
{\bar H_q}^2&\approx&\frac{1}{3}-\frac{(\bar\rho)^{\frac{1}{2}(\alpha-1)}}{2\sqrt{A}}\rightarrow\frac{1}{3},\nonumber\\
\sqrt{\kappa}\frac{d{\bar H}_q}{d\tilde t}&\approx&\frac{(\bar\rho)^{\frac{1}{2}(\alpha-1)}}{2\sqrt{A}}\rightarrow0,
\end{eqnarray}
and second and higher order derivatives of $\bar H_q$ with respect to $\tilde t$ also vanish when $a\rightarrow a_\textrm{max}$, i.e., $\bar\rho\rightarrow\infty$, because their leading order in the expansion on $\bar\rho$ is proportional to $(\bar\rho)^{(\alpha-1)/2}$ and $\alpha<-1$. Note that in this case $\tilde a=\sqrt{V}a\rightarrow+\infty$ and this happens at an infinite future $\tilde t$. Furthermore, we also find that
\begin{equation}
\tilde a\propto e^{H_q\tilde t}\textrm{ when }\tilde t\rightarrow\infty.
\end{equation}
Indeed, the Universe approaches a de Sitter state as described by the auxiliary metric in this case. Therefore, there is no singularity of the auxiliary metric when the auxiliary metric is on the form of a FLRW metric in the EiBI theory.
Note that $q^{\mu\nu}R_{\mu\nu}(\Gamma)\approx4/\kappa$ when $a\rightarrow a_\textrm{max}$ and $\tilde a\rightarrow\infty$ in this case.

\subsection{The EiBI scenario and the Type IV singularity}
\subsubsection{The physical metric $g_{\mu\nu}$}

To analyse the possibility of smoothing a type IV singularity within the EiBI theory, we consider the same kind of dark energy pGCG as shown in Eqs.~(\ref{pgcg EOM}) and (\ref{past density}), with $-1<\alpha<0$. Indeed, this fluid drives a past type IV singularity in GR except for some quantized cases, i.e., if $\alpha=-n/(n+1)$ with $n$ being natural numbers, the Hubble rate and all of its cosmic time derivatives are all regular in the finite past. Note that this result is different from the one proposed in Ref.~\cite{BouhmadiLopez:2007qb} because on that case the authors assumed a purely pGCG dominated universe for the analysis, which is not the case in this paper (see Eq.~\eqref{content}). {\color{black}See Ref.~\cite{Bouhmadi-Lopez:2014jfa} for more details on this issue.}

First, if $-1/2<\alpha<0$, the asymptotic behaviour of ${\bar H}^2$ and the derivatives of $\bar H$ as $\bar\rho_{de}\rightarrow0$, $a\rightarrow a_\textrm{min}$, $\bar\rho\rightarrow\bar\rho_r+\bar\rho_m\rightarrow\bar\rho_\textrm{ini}$, and $\bar p\rightarrow\bar\rho_r/3$ are
\begin{eqnarray}
{\bar H}^2&\approx&K(\bar\rho_{de})^{4\alpha+2},\nonumber\\
\frac{d^n}{d{\bar t}^n}\bar H&\propto&(\bar\rho_{de})^{1+(n+2)\alpha}+\textrm{higher order of }\bar\rho_{de},\nonumber\\
\label{typeiv}
\end{eqnarray}
where $n$ is a natural number and 
\begin{equation}
K=\frac{8\left[\bar\rho_\textrm{ini}+\bar\rho_r-2+2\sqrt{(1+\bar\rho_\textrm{ini})(1-\frac{1}{3}\bar\rho_r)^3}\right]}{27A^4\alpha^2(1+\bar\rho_\textrm{ini})(1-\frac{1}{3}\bar\rho_r)^{-2}},\nonumber
\end{equation}
where $\bar\rho_\textrm{ini}$ is the initial dimensionless energy density and $\bar\rho_r$ is the dimensionless radiation energy density evaluated at $a=a_\textrm{min}$. Furthermore, we can derive the asymptotic cosmic time behaviour near the singularity through the conservation equation Eq.~\eqref{conservation equation} to confirm that the Universe starts to expand from $a_{\textrm{min}}$ at a finite past cosmic time for this case. Actually, a universe will start from a finite past cosmic time as long as $-1<\alpha<0$ in the EiBI theory because 
\begin{equation}
\bar\rho_{de}\propto(\bar t-\bar t_\textrm{min})^{-\alpha}+\textrm{higher order of }(\bar t-\bar t_\textrm{min}),
\end{equation}
for $-1/2<\alpha<0$, and
\begin{equation}
\bar\rho_{de}\propto(\bar t-\bar t_\textrm{min})^{1+\alpha}+\textrm{higher order of }(\bar t-\bar t_\textrm{min}),
\end{equation}
for $-1<\alpha\le-1/2$ when $\bar\rho_{de}\rightarrow 0$ as well as $a\rightarrow a_\textrm{min}$.

According to Eqs.~\eqref{typeiv}, one can show that if $-1/2<\alpha<-1/3$, the first order cosmic time derivative of $\bar H$ goes to infinity and $\bar H$ is finite, implying a finite past sudden singularity.

If $\alpha=-1/(n+2)$ where $n$ is a positive integer (note that in this case $-1/3\le\alpha<0$), the Hubble rate and its higher order derivatives are all regular. The reason is the following: All the derivatives of the Hubble rate can be written as
\begin{equation}
\frac{d^n}{d{\bar t}^n}\bar H\propto D_n+(\bar\rho_{de})^{-\alpha}+\textrm{higher order of }\bar\rho_{de},
\end{equation}
with $D_n$ being a finite non-vanishing constant, which is finite at $a=a_\textrm{min}$ where $\bar\rho_{de}=0$. The next order derivative becomes
\begin{equation}
\frac{d^{n+1}}{d{\bar t}^{n+1}}\bar H\propto D_{n+1}+\textrm{higher order of }\bar\rho_{de},
\end{equation}
which still remains finite at $a=a_\textrm{min}$. We can then conclude that all the derivatives of $\bar H$ are regular at $a_\textrm{min}$ if $\alpha=-1/(n+2)$ with $n$ being a positive integer.

If, however, $-1/3\le\alpha<0$ and $\alpha\neq-1/(n+2)$, from Eq.~\eqref{typeiv} and the conservation equation we find that as long as $\alpha$ satisfies $-1/(p+2)<\alpha<-1/(p+3)$ with $p$ being a positive integer, the $(p+1)$-th derivative of $\bar H$ blows up while the $1$,...,$p$-th derivatives are all finite. This indicates a type IV singularity.

On the other hand, if $-1<\alpha\le-1/2$ and $\alpha$ cannot be written as $-n/(n+1)$ with $n$ being a natural number, we find that as long as $\alpha$ satisfies $-(p+1)/(p+2)<\alpha<-p/(p+1)$, the $p$-th derivative of $\bar H$ blows up while the $1$,...,$(p-1)$-th derivatives are all finite. This implies that a finite past type IV singularity except for a finite past sudden singularity in which the first order cosmic time derivative of the Hubble rate diverges if $-2/3<\alpha<-1/2$.

However, if $\alpha=-n/(n+1)$, the Hubble rate and its higher order derivatives are all regular. The reason is the following: All the derivatives of the Hubble rate can be written as 
\begin{equation}
\frac{d^n}{dt^n}\bar{H}\propto E_n+(\bar\rho_{de})^{1+\alpha}+\textrm{higher order of }\bar\rho_{de},
\end{equation}
with $E_n$ being a finite non-vanishing constant, are finite at $a=a_\textrm{min}$. The next order derivative hence becomes
\begin{equation}
\frac{d^{n+1}}{dt^{n+1}}\bar{H}\propto E_{n+1}+\textrm{higher order of }\bar\rho_{de},
\end{equation}
which still remains finite at $a=a_\textrm{min}$. We can then conclude that all the derivatives of $H$ are regular at $a_\textrm{min}$ if $\alpha=-n/(n+1)$ with $n$ being a positive integer.

In summary, we can summarize our results as follows:

\begin{itemize}
\item If $-1/2<\alpha<-1/3$ or $-2/3<\alpha<-1/2$, the Universe expands from a finite past sudden singularity.
\item If $-1/3<\alpha<0$ and $\alpha\neq-1/(n+2)$, or $-1<\alpha<-2/3$ and $\alpha\neq-n/(n+1)$, with $n$ being a positive integer, the Universe expands from a finite past type IV singularity.
\item If $\alpha=-1/(n+2)$ or $\alpha=-n/(n+1)$, there is no singularity and the Universe is born at a finite past.
\end{itemize}

On the above discussion, we have assumed that the total pressure $\bar p<1$ during the evolution of the Universe so that the left hand side of the modified field equation \eqref{field equation} is always positive. However, in some cases there may exist a particular scale factor $a_b$ satisfying $a_b>a_{\textrm{min}}$ where the total pressure $\bar p=1$ at $a_b$. Then, for this case the non-vanishing leading orders of the Hubble parameter and its cosmic time derivative in the expansion near $a_b$ are:
\begin{eqnarray}
{\bar H}^2&\propto&(\delta\bar p)^2,\nonumber\\
\frac{d\bar H}{d\bar t}&\propto&\delta\bar p,
\end{eqnarray}
where $\delta\bar p\equiv\bar p-1$. Hence, the Universe is born from a loitering effect in an infinite past, instead of the various past singularities mentioned previously.

\subsubsection{The auxiliary metric $q_{\mu\nu}$}
On the other hand, for $A>0$ it can also be shown that 
\begin{eqnarray}
{\bar H_q}^2&\approx&\frac{1}{3}+\frac{\bar\rho_\textrm{ini}+\bar\rho_r-2}{6\sqrt{(1+\bar\rho_\textrm{ini})(1-\frac{1}{3}\bar\rho_r)^3}},\nonumber\\
\sqrt{\kappa}\frac{d{\bar H}_q}{d\tilde t}&\approx&-\frac{\bar\rho_\textrm{ini}+\frac{1}{3}\bar\rho_r}{2\sqrt{(1+\bar\rho_\textrm{ini})(1-\frac{1}{3}\bar\rho_r)^3}},
\end{eqnarray}
near $a_\textrm{min}$ for $-1<\alpha<0$ and both the auxiliary Hubble rate and its first $\tilde{t}$ derivative are regular. Interestingly, we start from the conservation equation Eq.~\eqref{conservation equation} and find that the Universe starts from a finite $\tilde{t}$ as long as $-1<\alpha<0$. More precisely, we have
\begin{equation}
\bar\rho_{de}\propto(\tilde{t}-\tilde{t}_\textrm{min})^{1+\alpha}+\textrm{higher order of }(\tilde{t}-\tilde{t}_\textrm{min}),
\end{equation}
for $-1<\alpha<0$ when $\bar\rho_{de}\rightarrow 0$. This fact implies it is necessary to analyse the asymptotic behaviours of the higher order $\tilde{t}$ derivatives of the auxiliary Hubble rate to see whether there is a finite past type IV singularity of the auxiliary metric or not.

If $-1/2<\alpha<0$ and $\alpha$ cannot be written as $-1/(n+2)$ where $n$ is a positive integer, we find that as long as $\alpha$ satisfies $-1/(p+1)<\alpha<-1/(p+2)$, the $(p+1)$-th derivative of $\bar{H_q}$ blows up while the $1$,...,$p$-th derivatives are all finite. This indicates a type IV singularity of the auxiliary metric.

However, if $\alpha=-1/(n+2)$ where $n$ is a positive integer, the auxiliary Hubble rate and its higher order $\tilde{t}$ derivatives are all regular. The reason is the following: all the $\tilde{t}$ derivatives of the auxiliary Hubble rate can be written as 
\begin{equation}
\frac{d^n}{d\tilde{t}^n}\bar{H_q}\propto F_n+(\bar\rho_{de})^{-\alpha}+\textrm{higher order of }\bar\rho_{de},
\end{equation}
with $F_n$ being a finite non-vanishing constant, which are finite when $\bar{\rho}_{de}=0$. The next order derivative hence becomes
\begin{equation}
\frac{d^{n+1}}{d\tilde{t}^{n+1}}\bar{H_q}\propto F_{n+1}+\textrm{higher order of }\bar\rho_{de},
\end{equation}
which still remains finite. We can then conclude that all the $\tilde{t}$ derivatives of $H_q$ are regular as $\bar\rho_{de}\rightarrow0$ if $\alpha=-1/(n+2)$ with $n$ being a positive integer.

On the other hand, if $-1<\alpha\le-1/2$ and $\alpha$ cannot be written as $-n/(n+1)$ with $n$ being a natural number, we find that as long as $\alpha$ satisfies $-(p+1)/(p+2)<\alpha<-p/(p+1)$, the $(p+1)$-th derivative of $\bar{H_q}$ blows up while the $1$,...,$p$-th derivatives are all finite. This also implies that a past type IV singularity of the auxiliary metric.

However, if $\alpha=-n/(n+1)$, the auxiliary Hubble rate and its higher order $\tilde{t}$ derivatives are all regular. The reason is the following: all the $\tilde{t}$ derivatives of the auxiliary Hubble rate can be written on this case as 
\begin{equation}
\frac{d^n}{d\tilde{t}^n}\bar{H_q}\propto G_n+(\bar\rho_{de})^{1+\alpha}+\textrm{higher order of }\bar\rho_{de},
\end{equation}
with $G_n$ being a finite non-vanishing constant, which are finite when $\bar{\rho}_{de}=0$. The next order derivative hence becomes
\begin{equation}
\frac{d^{n+1}}{d\tilde{t}^{n+1}}\bar{H_q}\propto G_{n+1}+\textrm{higher order of }\bar\rho_{de},
\end{equation}
which still remains finite. We can then conclude that all the $\tilde{t}$ derivatives of $H_q$ are well defined as $\bar\rho_{de}\rightarrow0$ if $\alpha=-n/(n+1)$ with $n$ being a positive integer.

Thus, considering the auxiliary metric for $-1<\alpha<0$, the results can be summarized as follows:

\begin{itemize}
\item If $\alpha$ cannot be written as $-1/(n+2)$ or $-n/(n+1)$ with $n$ being a positive integer, the Universe expands from a type IV singularity of the auxiliary metric in which $\bar{H_q}$ and $d\bar{H_q}/d\tilde{t}$ are regular, while higher order $\tilde{t}$ derivatives of $H_q$ blow up at a finite $\tilde{t}$.
\item If $\alpha=-1/(n+2)$ or $\alpha=-n/(n+1)$, there is no singularity of the auxiliary metric and the Universe is born at a finite past $\tilde{t}$.
\end{itemize}

As for the case in which the singularities are replaced with a loitering effect of the physical metric ($\bar p\rightarrow 1$) discussed in the end of previous subsubsection, the loitering effect of the physical metric corresponds to a big bang singularity of the auxiliary metric compatible with the physical connection because both $H_q$ and $dH_q/d\tilde t$ diverge at a vanishing $\tilde a$ at a finite past $\tilde t$, and so does the Ricci scalar defined in Eq.~(\ref{aux}).

\subsection{The EiBI scenario and the Little Rip}
\subsubsection{The physical metric $g_{\mu\nu}$}

We conclude the analysis of this section by considering as well the possibility of smoothing a little rip event within the EiBI formalism. The Little Rip event is quite similar to the Big Rip singularity except that the former happens at an infinite future while the latter at a finite cosmic time. Such an event, despite avoiding a future singularity at a finite cosmic time, will still lead to the destruction of all structures in the Universe like the Big Rip. The Little Rip has been previously analysed under four-dimensional (4D) standard cosmology \cite{Ruzmaikina1970}, later on rediscovered on \cite{Nojiri:2005sx,Nojiri:2005sr,Stefancic:2004kb}. It can be found in dilatonic brane-world models  \cite{BouhmadiLopez:2005gk} or other kind of brane-world models \cite{Belkacemi:2011zk,Bouhmadi-Lopez:2013nma}. Forty years later after its discovery, the event has been {\textit{baptised}} and named the ``Little Rip'' \cite{Frampton:2011sp,Brevik:2011mm,Frampton:2011rh,Nojiri:2011kd}. 

The simplest and the most common-used dark energy equation of state driving the Little Rip in GR is \cite{Nojiri:2005sx,Stefancic:2004kb,Frampton:2011sp}
\begin{equation}
\bar p_{de}=-\bar\rho_{de}-B\sqrt{\bar\rho_{de}}, 
\label{eqstatelr}
\end{equation}
where $B$ is a positive dimensionless constant. After integrating the conservation equation (\ref{conservation equation}) of the dark energy fluid (\ref{eqstatelr}), one can easily check that its energy density $\bar\rho_{de}\rightarrow\infty$ as the scale factor $a\rightarrow\infty$. The asymptotic future behaviour of ${\bar H}^2$ and the cosmic time derivative of the Hubble rate as $a\rightarrow\infty$ are
\begin{eqnarray}
{\bar H}^2&\approx&\frac{\bar\rho}{3}\rightarrow\infty,\nonumber\\
\frac{d{\bar H}}{d\bar t}&\approx&\frac{B\sqrt{\bar\rho}}{2}=-\frac{\bar\rho+\bar p}{2}\rightarrow\infty.
\label{littleripHH}
\end{eqnarray}
Besides, for a finite $a_c$ (at a given cosmic time $\bar t_c$) very close to the Little Rip event (note that $a_c$ is large enough so that the asymptotic equations \eqref{littleripHH} are valid), the scale factor dependence on the cosmic time $\bar t$, can be approximated by
\begin{equation}
\frac{a}{a_c}\approx\exp\left\{\frac{2\sqrt{\bar\rho_{de_c}}}{3B}\left[e^{\frac{\sqrt{3}}{2}B(\bar t-\bar t_c)}-1\right]\right\},
\label{littletime}
\end{equation}
where $\bar\rho_{de_c}$ is the dimensionless dark energy density when $a=a_c$. As could be expected the radiation and dark matter components have no effect on the asymptotic behaviour and therefore where the Little Rip could take place.
Therefore, like in GR the scale factor, Hubble parameter and its cosmic time derivatives blow up in an infinite cosmic time where the Universe would hit a little rip. 

\subsubsection{The auxiliary metric $q_{\mu\nu}$}

Similarly it can be shown that for a matter content given by Eq.~\eqref{eqstatelr}, the asymptotic behaviours of ${\bar H_q}^2$ and $d\bar H_q/d\tilde t$ read
\begin{eqnarray}
{\bar H_q}^2&\approx&\frac{1}{3}-\frac{1}{3\bar\rho}+\frac{8+3B^2}{24(\bar\rho)^2}\rightarrow\frac{1}{3},\nonumber\\
\sqrt{\kappa}\frac{d{\bar H}_q}{d\tilde t}&\approx&\frac{B}{2}(\bar\rho)^{-\frac{3}{2}}\rightarrow0,\nonumber\\
\tilde t&\rightarrow&\infty,\nonumber\\
\tilde{a}&\propto&\tilde{t}e^{H_q\tilde t},
\label{littleripasy}
\end{eqnarray}
and other higher order $\tilde{t}$ derivatives of $\bar H_q$ approach zero when $\bar\rho\rightarrow\infty$.  Note again that $q^{\mu\nu}R_{\mu\nu}(\Gamma)\approx4/\kappa$ when $\bar\rho\rightarrow\infty$. Therefore, there is no Little Rip in the EiBI theory if one regards the auxiliary metric as the FLRW metric. Actually, the Universe approaches a de Sitter state as described by the scale factor \eqref{littleripasy}.

In summary, the asymptotic behaviours of the Universe filled with various kinds of dark energy, i.e., a phantom energy with a constant equation of state, a phantom energy driving a little rip event, and pGCG with different values of $\alpha$ on the basis of the EiBI theory are shown in TABLE.~\ref{summary}.

\begin{table*}
 \begin{center}
  \begin{tabular}{||c||c||c||}
  \hline 
   Singularity in GR & EiBI physical metric & EiBI auxiliary metric\\
  \hline\hline 
   Big Rip & Big Rip & expanding de-Sitter \\ 
  \hline\hline 
   past Sudden & past Type IV ($0<\alpha\leq2$) & contracting de-Sitter \\
   \cline{2-2}
   ($\alpha>0$)&past Sudden ($\alpha>2$)& \\
  \hline\hline 
   future Big Freeze &future Big Freeze ($-3<\alpha<-1$) & expanding de-Sitter \\
   \cline{2-2}
   ($\alpha<-1$)&future Type IV ($\alpha=-3$)&\\
   \cline{2-2}
   &future Sudden ($\alpha<-3$)&\\
  \hline\hline
   past Type IV&past Sudden ($-2/3<\alpha<-1/3$)&past Type IV\\
   \cline{2-2}
   ($-1<\alpha<0$)&(1)past Type IV&\\
   \cline{2-3}
   ($\alpha\neq-n/(n+1)$)&(2)finite past without singularity&finite past without singularity\\
   \cline{2-3}
   & past loitering effect ($a_b>a_\textrm{min}$)& Big Bang\\
  \hline\hline
  finite past without singularity&finite past without singularity&finite past without singularity\\
  ($\alpha=-n/(n+1)$)&&\\
  ($-1<\alpha<0$)&&\\
  \cline{2-3}
   & past loitering effect ($a_b>a_\textrm{min}$)& Big Bang\\
  \hline\hline
  Little Rip & Little Rip & expanding de-Sitter\\
  \hline
  \end{tabular}
  \caption{This table summarizes how the asymptotic behaviour of a universe near the singularities in GR is altered in the EiBI theory when the Universe is filled with matter, radiation as well as phantom energy. The row labelled by (1) corresponds to $-1/3<\alpha<0$ or $-1<\alpha<-2/3$, and where $\alpha$ cannot be written as $-1/(n+2)$ or $-n/(n+1)$, with $n$ being a natural number. If $\alpha=-1/(n+2)$ ($-1/3\le\alpha<0$ naturally) which is labelled by (2), there is no singularity while the Universe starts to expand from a finite size at a finite cosmic time. Note that it is possible for the Universe to start from a loitering phase of the physical metric instead of a past singularities, as long as the total pressure reaches the value $\bar p=1$ at some particular scale factor $a_b$ such that $a_b>a_{\textrm{min}}$, and it corresponds to a past big bang singularity of the auxiliary metric.}
    \label{summary}
 \end{center}
\end{table*} 

\section{The geodesic analyses of a Newtonian object within the EiBI setup}

In this section, we will consider a spherical Newtonian object with mass $M$ and a test particle rotating around the mass $M$ with a physical radius $r$. We assume that both of them are embedded in a spherically symmetric FLRW background. In Ref.~\cite{Faraoni:2007es}, the authors have shown that the bound systems with a strong enough coupling in a de-Sitter background will not comove with the accelerating expansion of the Universe. However it is not the case when general accelerating phases are considered, such as the various singularities we have analysed in this paper. Therefore, we will analyse the evolution equations of its physical radius, or the geodesic equations, when the Universe approaches those singularities. In the Palatini formalism, there are two metrics, the first one $g_{\mu\nu}$ couples to matter, the second one $q_{\mu\nu}$ is the auxiliary one which is compatible with the connection and fixes the curvature of the space-time. If we regard the first metric as the one used to define the distances, then the geodesic equation is then defined by the Levi-Civita connection of $g_{\mu\nu}$. On the other hand, if we consider the curvature, therefore $q_{\mu\nu}$, responsible for the geodesic equations then we can define another geodesic equation expressed by the coordinates $\tilde t$ and $\tilde a$ defined in Eq.~(\ref{defHq}). 

First, we regard the first metric $g_{\mu\nu}$ as the physical metric and the evolution equation of the physical radius reads \cite{Nesseris:2004uj,Faraoni:2007es}
\begin{equation}
\ddot{r}=\frac{\ddot{a}}{a}r-\frac{GM}{r^2}+\frac{L^2}{r^3},
\label{geodesic eq g}
\end{equation}
where the overdot denotes the cosmic time derivative and $L$ is the constant angular momentum per unit mass of the test particle. Essentially, $L$ satisfies the angular conservation equation \cite{Nesseris:2004uj,Faraoni:2007es}:
\begin{equation}
r^2\dot{\phi}=L
\label{angular momentum}
\end{equation}
in spherical coordinate.

 According to Ref.~\cite{Faraoni:2007es}, the $\ddot ar/a$ term can be treated as a perturbation when the object is embedded in the de-Sitter background. However, this is not the case as the Universe approaches the Big Rip, Little Rip, Big Freeze, and the Sudden singularities because of the divergence of $\ddot a/a$. In these cases, the evolution equation (\ref{geodesic eq g}) takes the form
\begin{equation}
\ddot{r}\approx\frac{\ddot{a}}{a}r,
\label{asy geodesic g}
\end{equation}
because the $\ddot ar/a$ dominates over the other terms in Eq.~(\ref{geodesic eq g}) \cite{Faraoni:2007es}. There are two solutions to Eq.~(\ref{asy geodesic g}) because it is a second order differential equation. One solution $r_1=a(t)$ is the \textit{trivial} solution, and the other solution can be derived directly through the following integration:
\begin{equation}
r_2=r_1\int\frac{dt}{{r_1}^2}.
\label{r2}
\end{equation}
Therefore, the solution to Eq.~(\ref{asy geodesic g}) is the linear combination of $r_1$ and $r_2$:
\begin{equation}
r(t)=A_1r_1(t)+A_2r_2(t).
\label{solution}
\end{equation}

Similarly, the angular motion of the particle can also be obtained by integrating Eq.~(\ref{angular momentum}), as shown in Ref.~\cite{Faraoni:2007es}.

\subsection{Dark energy with a constant equation of state: Big Rip case}
As the Universe approaches the Big Rip singularity in which the asymptotic behaviours of the Hubble rate and its cosmic time derivative take the form in Eqs~(\ref{HH in Big Rip}), the cosmic time dependence of the scale factor is similar to that in GR \cite{BouhmadiLopez:2004me,Faraoni:2007es}
\begin{equation}
a(t)\propto(t_\textrm{max}-t)^{\frac{2}{3(1+w)}}.
\end{equation}
Note that the exact analytical form of the previous equation was provided in Ref.~\cite{BouhmadiLopez:2004me}. Thus, we have the first trivial solution $r_1=a(t)$ and after integrating Eq.~(\ref{r2}) we can also derive the total solution
\begin{equation}
r(t)=A_1(t_\textrm{max}-t)^{\frac{2}{3(1+w)}}+A_2(t_\textrm{max}-t)^{1-\frac{2}{3(1+w)}}.
\label{br r}
\end{equation}
One can see that the evolution of the physical radius of the bound system is governed by the first solution in Eq.~(\ref{br r}) because the second one becomes negligible as the Big Rip is approached. Therefore, the physical radius of the object will comove and diverge with the scale factor.

Next, the angular motion can be obtained by integrating Eq.~(\ref{angular momentum}):
\begin{eqnarray}
\phi(t)&=&\int dt\frac{L}{r^2}\nonumber\\
&\approx&\frac{3(w+1)}{1-3w}\frac{L}{{A_1}^2}(t_\textrm{max}-t)^{\frac{3w-1}{3(w+1)}}+\phi_0,
\end{eqnarray}
where $\phi_0$ is a constant angle from now on.

Hence, one can see that $\phi(t)\rightarrow\phi_0$ as the Big Rip is approached, which means that the angular motion slows down and freezes near the singularity. These qualitative descriptions of the asymptotic behaviour of the bound system near the singularity confirm the existence of the Big Rip singularity in the EiBI theory.

\subsection{Phantom-GCG with $\alpha>2$: Sudden singularity case}
If the Universe approaches a finite past sudden singularity in which the asymptotic behaviours of the Hubble rate and its cosmic time derivative take the form in Eqs.~(\ref{HH sudden neq2}) for $\alpha>2$ (for the sake of convenience, we will only consider $\alpha>2$ even if there are other regions of the parameter space in which the Sudden singularity occurs), the cosmic time dependence of the scale factor which can be derived from Eq.~(\ref{cosmic time neq 2}) is
\begin{equation}
\frac{a(t)}{a_\textrm{min}}\propto1+C_{S}(t-t_\textrm{min})^{\frac{4(1+\alpha)}{3\alpha}},
\end{equation}
where $C_{S}$ is a positive constant. Following a similar procedure as in the previous subsection, we have the first trivial solution $r_1=a(t)$ and after integrating Eq.~(\ref{r2}) we can also derive the total solution
\begin{eqnarray}
r(t)&\approx&A_1\left[1+C_{S}(t-t_\textrm{min})^{\frac{4(1+\alpha)}{3\alpha}}\right]\nonumber\\
&+&A_2(t-t_\textrm{min}).
\label{s r}
\end{eqnarray}

One can see that the evolution of the physical radius of the bound system is also governed by the first solution in Eq.~(\ref{s r}) because the second one becomes negligible as $t\rightarrow t_\textrm{min}$:
\begin{equation}
r(t)\approx A_1,
\end{equation}
which can be shown to be similar to the behaviour near the Sudden singularity in GR \cite{BouhmadiLopez:2007qb}. 

On the other hand, the angular motion in this case is
\begin{equation}
\dot{\phi}\approx\frac{L}{{A_1}^2},\nonumber
\end{equation}
therefore,
\begin{equation}
\phi(t)\approx\frac{L}{{A_1}^2}(t-t_\textrm{min})+\phi_0.
\end{equation}
Thus, the particle starts its motion from $r(t_\textrm{min})=A_1$, $\phi(t_\textrm{min})=\phi_0$, with an infinite radial acceleration $\ddot{r}$ at the past singularities.

\subsection{Phantom-GCG with $-3<\alpha<-1$: Big Freeze case}
If the Universe approaches a finite future big freeze singularity in which the asymptotic behaviours of the Hubble parameter and its cosmic time derivative take the form in Eqs.~(\ref{big freeze beq -3}) for $-3<\alpha<-1$ (for the sake of convenience, we will only consider $-3<\alpha<-1$ even if there are other regions of the parameter space in which the Big Freeze singularity occurs), the cosmic time dependence of the scale factor which can be derived from Eq.~(\ref{alpha not -3 t}) is
\begin{equation}
\frac{a(t)}{a_\textrm{max}}\propto1-C_{BF}(t_\textrm{max}-t)^{\frac{4(1+\alpha)}{1+3\alpha}},
\end{equation}
where $C_{BF}$ is a positive constant. Thus, we have the first trivial solution $r_1=a(t)$ and after integrating Eq.~(\ref{r2}) we can also derive the total solution
\begin{eqnarray}
r(t)&\approx&A_1\left[1-C_{BF}(t_\textrm{max}-t)^{\frac{4(1+\alpha)}{1+3\alpha}}\right]\nonumber\\
&+&A_2(t_\textrm{max}-t).
\label{bf r}
\end{eqnarray}

One can see that the evolution of the physical radius of the bound system is also governed by the first solution in Eq.~(\ref{bf r}) because the second one becomes negligible as $t\rightarrow t_\textrm{max}$:
\begin{equation}
r(t)\approx A_1,
\end{equation}
which can be shown to be similar to the behaviour near the Big Freeze in GR \cite{BouhmadiLopez:2007qb}.

Similarly, the angular motion of this particle near the singularities is
\begin{equation}
\dot{\phi}\approx\frac{L}{{A_1}^2},\nonumber
\end{equation}
consequently,
\begin{equation}
\phi(t)\approx-\frac{L}{{A_1}^2}(t_\textrm{max}-t)+\phi_0.
\label{angular bf}
\end{equation}
Thus, the particle will remain its bound structure at $r(t_\textrm{max})=A_1$ and $\phi(t_\textrm{max})=\phi_0$, with an infinite radial acceleration $\ddot{r}$ at the future singularity.

\subsection{Dark energy driving the Little Rip event}
If the Universe approaches a little rip singularity in which the asymptotic behaviours of the Hubble rate and its cosmic time derivative take the form in Eqs.~(\ref{littleripHH}), the cosmic time dependence of the scale factor which can be derived from Eq.~(\ref{littletime}) is
\begin{equation}
a(t)\approx a_c\exp\left\{\frac{2\sqrt{\bar\rho_{de_c}}}{3B}\left[e^{\frac{\sqrt{3}}{2}B(\bar t-\bar t_c)}-1\right]\right\},
\end{equation}
where $a_c$ is the scale factor when the Universe is close enough to the Little Rip. Thus, we have the first trivial solution $r_1=a(t)$ and after integrating Eq.~(\ref{r2}) we can also derive the total solution
\begin{eqnarray}
r(t)&\approx&A_1\exp\left\{\frac{2\sqrt{\bar\rho_{de_c}}}{3B}\left[e^{\frac{\sqrt{3}}{2}B(\bar t-\bar t_c)}-1\right]\right\}\nonumber\\
&+&A_2\exp\left\{\frac{2\sqrt{\bar\rho_{de_c}}}{3B}\left[e^{\frac{\sqrt{3}}{2}B(\bar t-\bar t_c)}+1\right]\right\}\nonumber\\
&\times& Ei\left[-\frac{4\sqrt{\bar\rho_{de_c}}}{3B}e^{\frac{\sqrt{3}}{2}B(\bar t-\bar t_c)}\right],
\label{li r}
\end{eqnarray}
where $Ei[z]$ is the exponential integral functions \cite{abram}. We have numerically found that the second term in Eq.~\eqref{li r} proportional to $A_2$ vanishes near the Little Rip event, that is, when $\bar t\rightarrow\infty$. Thus, the evolution of the physical radius of the bound system is also governed by the first solution in Eq.~(\ref{li r}) 

Similarly, the angular motion of this particle near the Little Rip is
\begin{equation}
\dot{\phi}\approx\frac{L}{a^2},\nonumber
\end{equation}
consequently,
\begin{equation}
\phi(t)\approx\phi_0.
\label{angular bf}
\end{equation}
Therefore, one can see that the angular motion slows down and freezes near the Little Rip event while the radius of the bound system blows up near the Little Rip event. These qualitative descriptions of the asymptotic behaviour of the bound system near the singularity confirm the existence of the Little Rip event in the EiBI theory.

{\color{black}Furthermore, following a similar approach it can be seen that the bound structures will not be destroyed near the Type IV singularity because the singularity is too weak and therefore $\ddot{a}/a$ will always be finite. For the auxiliary metric we got a similar results as for the Type IV singularity within the physical metric.}

\begin{widetext}

\begin{figure}[h!]
\begin{center}
\includegraphics[scale=0.25]{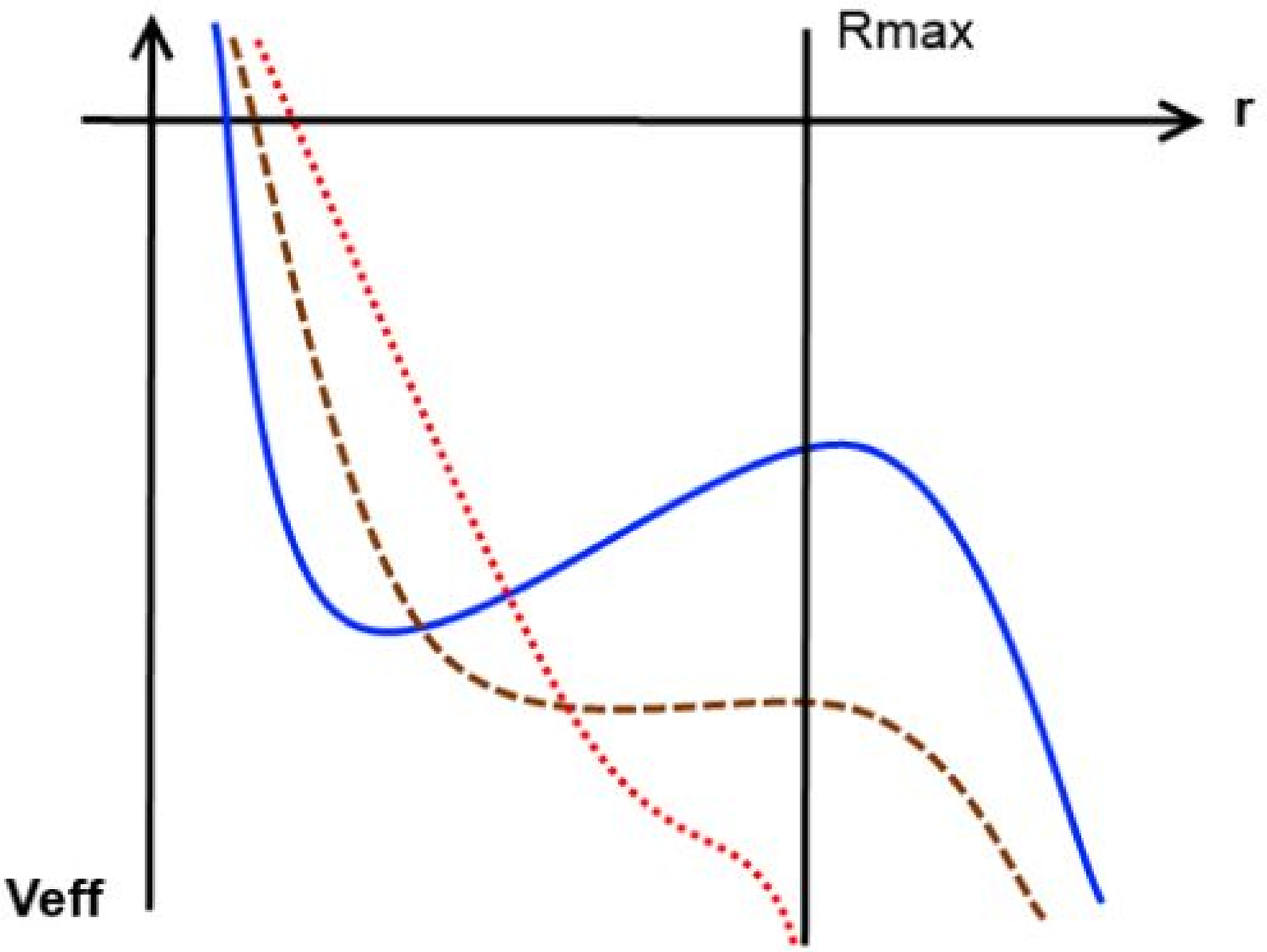} \hspace*{0.5cm}
\includegraphics[scale=0.25]{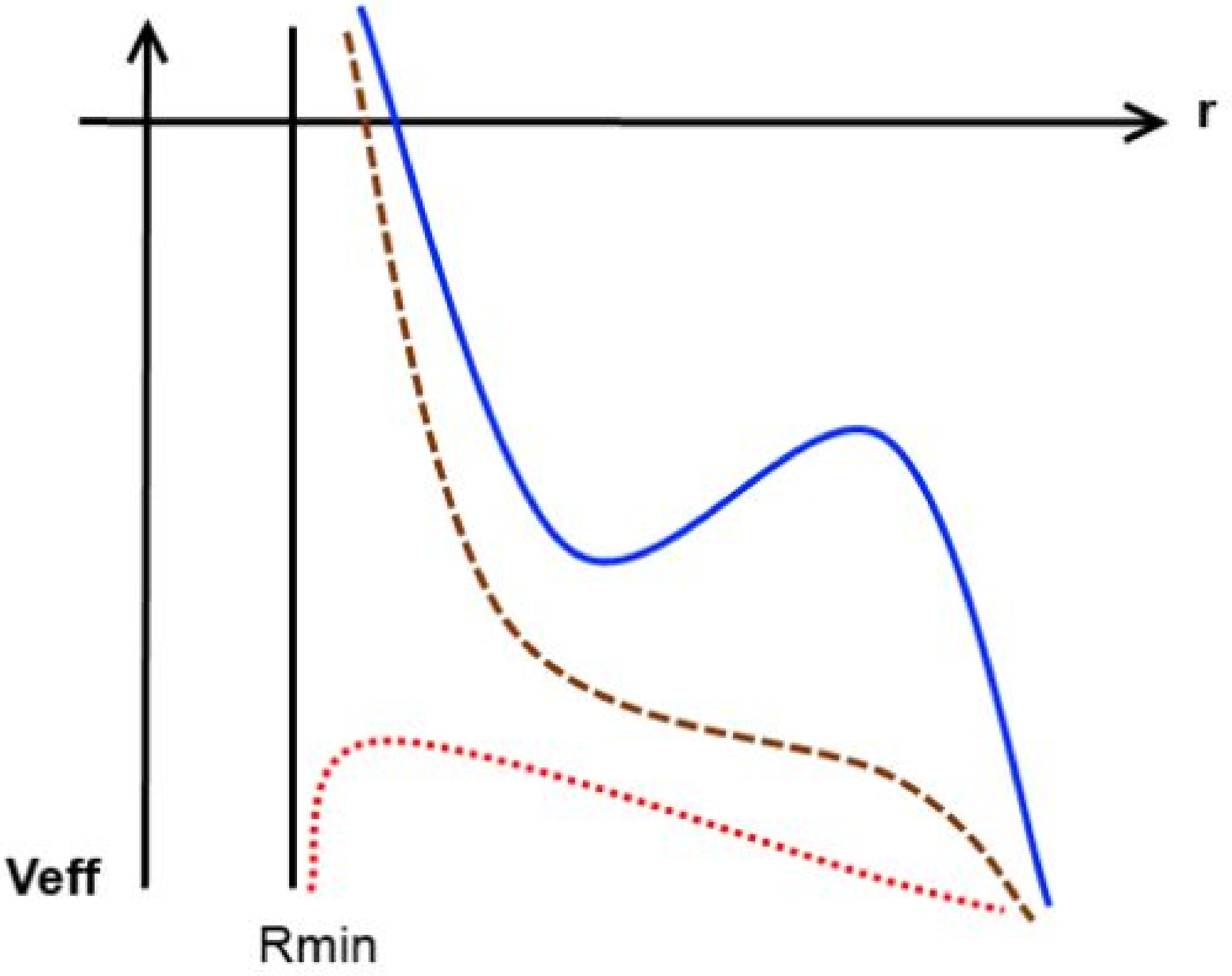}
\caption{We show the behaviour of the effective potential (\ref{Veff}) for future singularities (left figure) and past singularities (right figure). $R_{\textrm{max}}$ is finite for a sudden and big freeze singularities while infinite for a big rip and little rip singularity. Likewise $R_{\textrm{min}}$ is finite for a past sudden singularity. On the left figure: the blue solid curve shows the current bound structure, the brown dashed one the intermediate future behaviour and the red dotted one the final state. On the right figure the colors appear in an inverted chronological order, first red dotted, then brown dashed and finally blue solid, as the singularity takes place in the past.}
\label{Veffplot}
\end{center}
\end{figure}

\end{widetext}

Before concluding this section, we would like to mention briefly an alternative method to analyse the fate of bound structure. More precisely, the motion of a test particle moving around a massive object of mass $M$ is described by the equation of motion (\ref{geodesic eq g}), or alternatively one can invoke the effective potential \cite{Nesseris:2004uj}
\begin{equation}
V_{\textrm{eff}}=-\frac12 \frac{\ddot a}{a}r^2-\frac{GM}{r}+\frac12\frac{L^2}{r^2}
\label{Veff}
\end{equation}
where $\dot{r}^2=-2V_{\textrm{eff}}$. The existence of a bound structure with a circular orbit around the massive body $M$ corresponds to the existence of a minimum of the potential $V_{\textrm{eff}}$. We schematically show the time evolution of our effective potential in FIG.~\ref{Veffplot} which of course confirm our previous approximated results based on the geodesic equation (\ref{geodesic eq g}).

{\color{black} We would like to stress that the  analysis we have performed and which is based on the evolution equation \eqref{geodesic eq g} is valid for  Newtonian objects and under a weak field limit. Actually, there are many choices of metrics one can use to interpolate between a Schwarzschild and a FLRW metric (see Ref.~\cite{Baker:2001yc}). Our results may be improved if a better interpolating metric is chosen. However, we expect the approximation in Eq.~\eqref{asy geodesic g} is still valid as the Universe is close to the cosmological singularities considered above because the matter parts (the terms proportional to $GM$ in the evolution equation of the bound structure) are small compared with the expansion terms within such situations. Therefore, the use of the evolution equation 
\eqref{geodesic eq g} is quite fair in our analysis. Furthermore, we did not focus so much on the kind of gravitating systems we are considering but on the end state of the gravitating system in an expanding FLRW background of the kind we have analysed on the previous section. We have also followed a GR approach on this analysis because (i) for the strongest singularity like the Big Rip, the EiBI theory would behave at first order as GR with a different gravitational constant and (ii) for simplicity, we can improve the interpolating metric between a Schwarzschild and a FLRW metric but then we will need to take into account the gravitational theory we are analysing. We think this is far beyond the scope of this paper and we will come back to this issue in the future. Finally, we would like to finish by noticing that even for the strongly gravitating system such as the case of a black hole, the approach we have followed or a more exhaustive one as the one presented in \cite{Faraoni:2007es} lead to the same result: the black hole event horizon is destroyed. If results are consistent for strongly gravitating system we see no reasons why the results will be modified for other kind of systems. Of course all these hold in GR and we expect it is still valid on the EiBI theory for the reasons stated above.}

\section{A cosmographic approach of the EiBI scenario}

In this section, we will use the cosmographic approach to constrain the parameters of our model, especially in the cases in which the singularities are driven by the pGCG introduced in the previous sections. The cosmographic approach does not assume any particular form of the Friedmann equations and only depends on the assumption that the space-time is described by a FLRW metric. This makes this approach completely model independent so that we can use it to constrain the parameters of the pGCG model in the EiBI framework \cite{Capozziello:2008qc,Capozziello:2009ka,BouhmadiLopez:2010pp,Capozziello:2011tj}. In Section II, one can see that the parameters in this theory have a profound influence on the doomsday or the birth of the Universe.  Therefore, if the parameters in this theory are somehow constrained, one can further forecast the future evolution of the Universe and the possibility of past singularities different from the Big Bang.

The starting point of the cosmographic approach is the Taylor expansion of the scale factor $a(t)$ with respect to the cosmic time $t$ around the present time $t_0$ \cite{Capozziello:2008qc,Capozziello:2009ka,BouhmadiLopez:2010pp,Capozziello:2011tj}:
\begin{equation}
a(t)\equiv1+\sum^\infty_{i=1}\frac{1}{i!}\frac{d^ia}{dt^i}|_{t=t_0}(t-t_0)^i.
\end{equation}
It is convenient to define the following cosmographic parameters:
\begin{eqnarray}
H(t)&=&\frac{1}{a}\frac{da}{dt},\nonumber\\
q(t)&=&-\frac{1}{a}\frac{d^2a}{dt^2}\frac{1}{H^2},\nonumber\\
j(t)&=&\frac{1}{a}\frac{d^3a}{dt^3}\frac{1}{H^3},\nonumber\\
s(t)&=&\frac{1}{a}\frac{d^4a}{dt^4}\frac{1}{H^4},\nonumber\\
l(t)&=&\frac{1}{a}\frac{d^5a}{dt^5}\frac{1}{H^5},
\label{cosp}
\end{eqnarray}
which are commonly called the Hubble, deceleration, jerk, snap and lerk parameters \cite{Capozziello:2008qc,Capozziello:2009ka,BouhmadiLopez:2010pp,Capozziello:2011tj}. Furthermore, one can use the definitions in Eqs.~(\ref{cosp}) to derive the relations between these parameters and the redshift $z$ derivatives of the square of the Hubble rate \cite{Capozziello:2011tj}:
\begin{eqnarray}
\frac{d(H^2)}{dz}&=&\frac{2H^2}{1+z}(1+q),\nonumber\\
\frac{d^2(H^2)}{dz^2}&=&\frac{2H^2}{(1+z)^2}(1+2q+j),\nonumber\\
\frac{d^3(H^2)}{dz^3}&=&\frac{2H^2}{(1+z)^3}(-qj-s),\nonumber\\
\frac{d^4(H^2)}{dz^4}&=&\frac{2H^2}{(1+z)^4}\nonumber\\
&\times&(4qj+3qs+3q^2j-j^2+4s+l).
\end{eqnarray}
Next, one can evaluate these quantities at the present time:
\begin{eqnarray}
\frac{d(H^2)}{dz}|_{z=0}&=&2{H_0}^2(1+q_0),\nonumber\\
\frac{d^2(H^2)}{dz^2}|_{z=0}&=&2{H_0}^2(1+2q_0+j_0),\nonumber\\
\frac{d^3(H^2)}{dz^3}|_{z=0}&=&2{H_0}^2(-q_0j_0-s_0),\nonumber\\
\frac{d^4(H^2)}{dz^4}|_{z=0}&=&2{H_0}^2\nonumber\\
&\times&(4q_0j_0+3q_0s_0+3{q_0}^2j_0-{j_0}^2+4s_0+l_0),\nonumber\\
\label{coupled equations}
\end{eqnarray}
where the subscript $0$ denotes the quantities at the present time.

With the above equations and definitions, we can basically use the matter content given in Eq.~(\ref{content}), regarding the pGCG as the dark energy component, and rewrite the modified Friedmann equation (\ref{field equation}) as a function of the redshift $z$ then taking its $z$ derivatives. There are six parameters in our model: $\kappa$, $\alpha$, $a_{\textrm{max}}$ (or $a_{\textrm{min}}$), $\Omega_m$, $\Omega_{de}$, and $\Omega_r$ where the last three are the density parameters of dark and baryonic matter, dark energy, and radiation, respectively. For the remainder of this paper, we will assume $\Omega_r=8.48\times 10^{-5}$ according to Ref.~\cite{modern cos}. Therefore, we are left with five parameters and we can in principle use Eqs.~(\ref{coupled equations}) and $(H/H_0)^2|_{z=0}=1$ to close our system and constrain our model as long as all the cosmographic parameters are given. However, one has to keep in mind that the past evolution of the Universe has imposed some physical constraints on the parameters of the model. For example, when one considers the past singularities, i.e., $\alpha>-1$, the minimum scale factor $a_{\textrm{min}}$ should be very small to make this model in accordance with the well-known evolution of the Universe. With this assumption, one may expect that these cases should be very close to the $\Lambda$CDM version of the EiBI theory as the dark energy density approaches a constant at the present time (see Eq.~\eqref{past density}). On the other hand, when one considers the future Big Freeze singularities, these physical restrictions are loosened. 

In the following analysis, we will use two different methods to constrain our model, depending on which kind of singularity we analyse: (1) we will define a new dimensionless parameter 
\begin{equation}
Y\equiv\frac{x}{1-j_0+2\Omega_r},
\label{defineY}
\end{equation}
in which $x\equiv {a_s}^{3(1+\alpha)}$ where $a_s$ corresponds to the location of the singularity; i.e., it corresponds to $a_{\textrm{min}}$ or $a_{\textrm{max}}$ depending on the value of $\alpha$, then we leave $Y$ as a free variable for the sake of convenience of the computations. (2) we can also assume that $\Omega_m$ is model independent, that is, $\Omega_m=0.315$ according to the Planck mission \cite{CMB} when the models in which future singularities occur are dealt with, because there is no physical constraint on the maximum scale factor $a_{\textrm{max}}$. Furthermore, we can assume in both approaches $\Omega_\kappa\equiv3\kappa {H_0}^2$ is very small according to the results in Refs.~\cite{Avelino:2012ge,Avelino:2012qe}, where the authors showed that $\Omega_\kappa$ is much smaller than the other density parameters $\Omega_m$ and $\Omega_{de}$. In this way, we only need two cosmographic parameters $q_0$ and $j_0$ to constrain our model and we can avoid suffering from the large error bars when other cosmographic parameters $s_0$ and $l_0$ are taken into account \cite{Gruber:2013wua,Capozziello:2011tj}.

In summary, our strategy will be the following: (i) method A: Even though we have five free parameters (note that $\Omega_r$ has been fixed), $\Omega_\kappa$ can be chosen as a small number \cite{Avelino:2012ge,Avelino:2012qe}. Therefore, we are left with only four free parameters which are constrained by the Friedmann equation $H/H_0$, $q_0$ and $j_0$, leaving $Y$ as the free parameter. (ii) method B: Again, and even though, we have five parameters, $\Omega_\kappa$ can be chosen as a small number \cite{Avelino:2012ge,Avelino:2012qe} and $\Omega_m$ can be fixed by Planck data. Therefore, we are left again with only three free parameters which are constrained by the Friedmann equation $H/H_0$, $q_0$ and $j_0$.

Note that the cosmographic approach is a kinematic approach very useful when combined with the observational data of the current universe. In addition, the EiBI theory is very close to GR because $\Omega_\kappa$ is very small \cite{Avelino:2012ge,Avelino:2012qe}. In GR, one can derive the following relations from Eqs.~\eqref{rpp}, \eqref{rfs}, and Eqs.~\eqref{coupled equations}:
\begin{eqnarray}
1&=&\Omega_r+\Omega_m+\Omega_{de},\nonumber\\
2+2q_0&=&4\Omega_r+3\Omega_m-3\Omega_{de}X,\nonumber\\
2+4q_0+2j_0&=&12\Omega_r+6\Omega_m-3\Omega_{de}(3\alpha+2)X\nonumber\\
&-&9\Omega_{de}\alpha X^2,
\label{gr1}
\end{eqnarray}
where $X\equiv x/(1-x)$ and $0<x= {a_s}^{3(1+\alpha)}<1$.

If we insert the dimensionless parameter $Y=x/(1-j_0+2\Omega_r)$ defined previously and keep it as a free parameter whose value changes between $0$ and $1/(1-j_0+2\Omega_r)$, in GR we can further express $\alpha$, $a_s$, and $\Omega_m$ as functions of $Y$, $q_0$, and $j_0$ analytically:
\begin{eqnarray}
\alpha&=&\frac{2[1-(1-j_0+2\Omega_r)Y]}{3(1-2q_0+\Omega_r)Y},\label{alphaGR}\\
a_s&=&\Big[(1-j_0+2\Omega_r)Y\Big]^{\frac{1}{3(1+\alpha)}}\nonumber\\
&=&\Big[(1-j_0+2\Omega_r)Y\Big]^{\frac{(1-2q_0+\Omega_r)Y}{2-2(1-j_0)Y+3(1-2q_0)Y-\Omega_r Y}},\label{amGR}\\
\Omega_m&=&1-\Omega_r-\frac{(1-2q_0+\Omega_r)[1-(1-j_0+2\Omega_r)Y]}{3}.\nonumber\\\label{omGR}
\end{eqnarray}
Note as well that $Y$ and $1-j_0+2\Omega_r$ have the same sign because $0<Y(1-j_0+2\Omega_r)=x<1$. Before concluding, we would like to stress that we have $4$ parameters on the GR setup: $\alpha$, $a_s$, $\Omega_m$ and $\Omega_{de}$ and three constraints: the Friedmann equation evaluated at present, the observational values of $q_0$ and $j_0$. Therefore we are left with a unique degree of freedom or free parameter that we have chosen as $Y$.

Before solving numerically the cosmographic constraints in the EiBI theory, we will provide some qualitative behaviours of $\alpha$, $a_s$ and $\Omega_m$ as functions of $Y$ in GR (see Eqs.~(\ref{alphaGR}), (\ref{amGR}), and (\ref{omGR})). First of all, one can see from Eq.~\eqref{alphaGR} that $\alpha\rightarrow+\infty$ $(-\infty)$ for $Y\rightarrow0^{+}$ $(0^{-})$ if $1-j_0+2\Omega_r$ is positive (negative), and $\alpha\rightarrow0$ for $Y\rightarrow1/(1-j_0+2\Omega_r)$. Note that $q_0$ is always negative in an accelerating universe as it is in our case. Second, from Eq.~\eqref{amGR} one can see that $a_s\rightarrow1$ for the limits $Y\rightarrow0$ and $Y\rightarrow1/(1-j_0+2\Omega_r)$. Note that the right hand side of Eq.~\eqref{amGR} is always smaller than $1$ if $Y$ and $\alpha$ are positive, corresponding therefore $a_s$ to $a_{\textrm{min}}$ (See the bottom figure in FIG.~\ref{alphaam999}). For negative $Y$ and $\alpha$, the values of $a_s$ defined in Eq.~\eqref{amGR} can be divided into $a_{\textrm{min}}$ and $a_{\textrm{max}}$ by a particular $Y$ whose absolute value reads $|Y_p|$, which corresponds to $1+\alpha=0$. Besides, we find that $a_{\textrm{min}}$ has a local minimum for $1-j_0+2\Omega_r>0$. Furthermore, there is a positive, divergent $a_{\textrm{max}}$ at $|Y_p|^{-}$ corresponding to $1+\alpha\rightarrow0^{-}$ and a vanishing $a_{\textrm{min}}$ at $|Y_p|^{+}$ corresponding to $1+\alpha\rightarrow0^{+}$ for $1-j_0+2\Omega_r<0$. Finally, one can see from Eq.~\eqref{omGR} that $\Omega_m$ is a straight line ranging from $(2+2q_0-4\Omega_r)/3$ $(Y\rightarrow0)$, which corresponds exactly to $\Omega_m$ in the radiation$+\Lambda$CDM model, to $1-\Omega_r$ $(Y\rightarrow1/(1-j_0+2\Omega_r))$, which is exactly a pure radiation$+$CDM model.

In the following subsections, we will apply the two approaches enumerated previously just after Eqs.~(\ref{coupled equations}).

\subsection{The first method: introducing $Y$}

The most recent cosmographic analysis based on SNeIa observational data has been carried out in Ref.~\cite{Gruber:2013wua} (as far as we know)~\footnote{We will use the cosmographic results obtained in Ref.~\cite{Gruber:2013wua} but please notice that for the purpose of the current work we could have taken other works from the one available on the literature.}. The authors amended the conventional methodology of cosmography employing Taylor expansions of observables by an alternative method using Pad$\acute{\textrm{e}}$ approximations, and claimed that the numerical fitting analysis for the cosmographic parameters is improved substantially by this mean. Their analysis is based on Type Ia supernovae data from the Union 2.1 compilation of the supernova cosmology project. They performed several fits distinguished by numbers $(1)$ to $(7)$ \cite{Gruber:2013wua}: 
\begin{itemize}
\item Fit $(1)$: The analysis using the Taylor approach without priors. ($H_0=69.90^{+0.438}_{-0.433}$, $q_0=-0.528^{+0.092}_{-0.088}$, $j_0=0.506^{+0.489}_{-0.428}$)
\item Fit $(2)$: The analysis using the Pad$\acute{\textrm{e}}$ parametrization without priors. ($H_0=70.25^{+0.410}_{-0.403}$, $q_0=-0.683^{+0.084}_{-0.105}$, $j_0=2.044^{+1.002}_{-0.705}$)
\item Fit $(3)$: The analysis using the Pad$\acute{\textrm{e}}$ parametrization with the short redshift range $z\in[0,0.36]$. ($H_0=70.090^{+0.460}_{-0.450}$, $q_0=-0.658^{+0.098}_{-0.098}$, $j_0=2.412^{+1.065}_{-0.978}$)
\item Fit $(4)$: The analysis presuming priors from Planck's results on $H_0$ only. ($H_0=67.11$, $q_0=-0.069^{+0.051}_{-0.055}$, $j_0=-0.955^{+0.228}_{-0.175}$)
\item Fit $(5)$: The analysis presuming priors from Planck's results on $q_0$ only. ($H_0=69.77^{+0.288}_{-0.290}$, $q_0=-0.513$, $j_0=-0.785^{+0.220}_{-0.208}$)
\item Fit $(6)$: The analysis presuming priors from Planck's results on both $H_0$ and $q_0$. ($H_0=67.11$, $q_0=-0.513$, $j_0=2.227^{+0.245}_{-0.237}$)
\item Fit $(7)$: The analysis presuming priors on $H_0$ from the first-order fit of the luminosity distance. ($H_0=69.96^{+1.12}_{-1.16}$, $q_0=-0.561^{+0.055}_{-0.042}$, $j_0=0.999^{+0.346}_{-0.468}$)
\end{itemize}
Besides, the authors also showed that fits $(2)$, $(3)$ and $(7)$ seem to have the most reasonable results. Note that the results in fit $(7)$ are nearly identical to the $\Lambda$CDM model (see TABLE I of Ref.~\cite{Gruber:2013wua}) because in the $\Lambda$CDM model, $q_0=-1+3\Omega_m/2$ and $j_0=1$, so we will be mainly using fit $(7)$ in this subsection, especially for the cases in which past singularities could happen.

\subsubsection{The analyses for positive $Y$ in the EiBI theory}

\begin{figure}[!h]
\begin{center}
\includegraphics[scale=0.8]{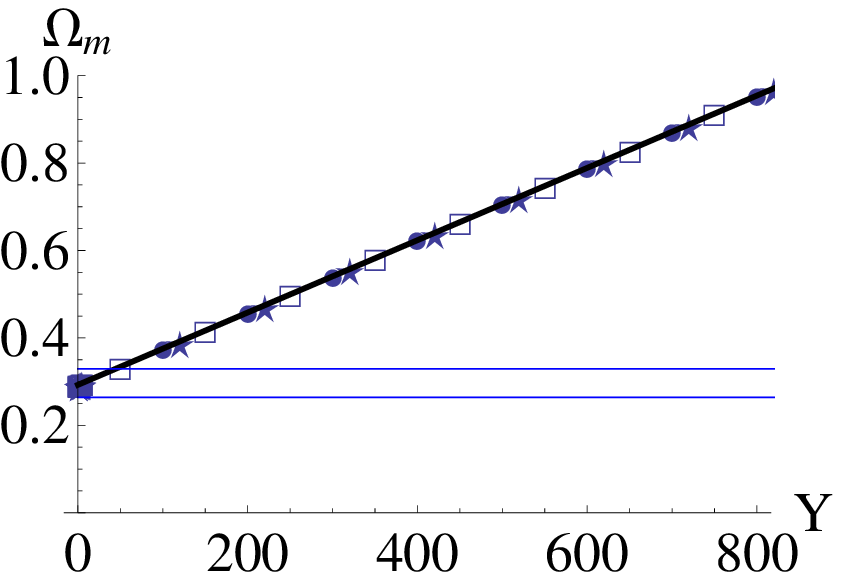}
\includegraphics[scale=0.8]{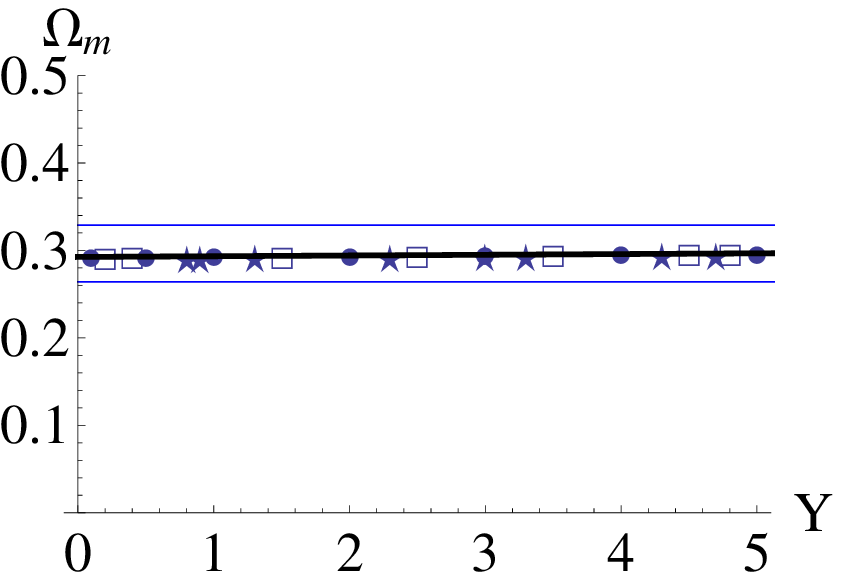}
\caption{The numerical results of $\Omega_m$ derived with fit $(7)$ in \cite{Gruber:2013wua} in which $q_0=-0.561$, $j_0=0.999$ and $\Omega_r=8.48\times 10^{-5}$ for positive $Y$. On the top figure $Y$ changes from $0$ to $1/(1-j_0+2\Omega_r)$, and on the bottom one $Y$ changes from $0$ to $5$. The circle, block, star symbols, and the black solid line correspond to the numerical results by imposing $\Omega_\kappa=10^{-6}$, $10^{-7}$, $10^{-8}$, and the analytical result in GR described by Eq.~(\ref{omGR}) ($\Omega_\kappa=0$), respectively. The region between the horizontal blue lines represents the $1\sigma$ errors of $\Omega_m$ derived from the given $q_0$ on the basis of the $\Lambda$CDM model ($j_0=1$ and $\Omega_r=0$).}
\label{om999}
\end{center}
\end{figure}

First, we use fit $(7)$ of Ref.~\cite{Gruber:2013wua} in which $q_0=-0.561$, and $j_0=0.999$ to evaluate $\Omega_m$ for different $Y$ and $\Omega_\kappa$ in the EiBI theory. According to the definition of $Y$ given in Eq.~(\ref{defineY}), only positive $Y$ need to be considered here because $1-j_0+2\Omega_r$ is positive. The results are shown in FIG.~{\ref{om999}}. The circle, block, star symbols, and the black solid line correspond to the numerical results imposing $\Omega_\kappa=10^{-6}$, $10^{-7}$, $10^{-8}$, and the analytical GR result described by Eq.~\eqref{omGR} ($\Omega_\kappa=0$), respectively. Additionally, we also include the $1\sigma$ errors of $\Omega_m$ from the constraint of $q_0$ in this fit on the basis of the $\Lambda$CDM model, which is shown in the region between the blue lines.

From these two figures, one can obtain two simple conclusions: (i) We do not see much difference between using EiBI and GR, the reason of course is that $\Omega_\kappa$ is very small as predicted in Refs.~\cite{Avelino:2012ge,Avelino:2012qe}. (ii) In order to obtain values of $\Omega_m$ compatible with the $\Lambda$CDM model, which we will consider as a guiding line of our analysis, we will stick to small values of $Y$ which we will consider to be smaller than $5$.

\begin{figure}[!h]
\begin{center}
\includegraphics[scale=0.8]{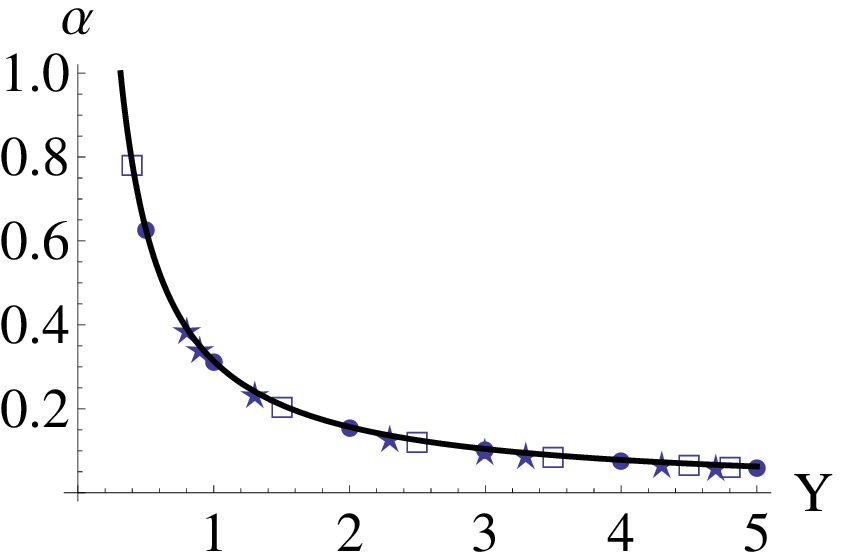}
\includegraphics[scale=0.8]{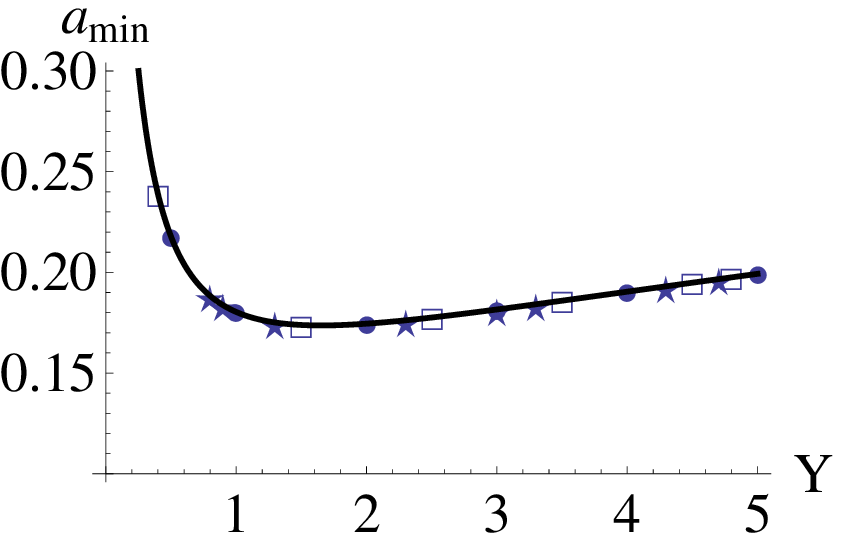}
\caption{The numerical constraints on $\alpha$ (top) and $a_{\textrm{min}}$ (bottom) versus positive $Y$. The points in different symbols denote the numerical results in different $\Omega_\kappa$ with the same imposed values in FIG.~{\ref{om999}}, and the black solid line corresponds to the analytical curves in GR, that is, Eqs.~\eqref{alphaGR} and \eqref{amGR}.}
\label{alphaam999}
\end{center}
\end{figure}

In FIG.~\ref{alphaam999}, one can see again that the EiBI theory does not make any distinguishable difference on the results of $\alpha$ and $a_{\textrm{min}}$ as functions of $Y$ from GR. Besides, though $a_{\textrm{min}}$ approaches $1$ both at $Y\rightarrow0$ and $Y\rightarrow1/(1-j_0+2\Omega_r)$ ($\alpha\rightarrow\infty$ and $0$, respectively), there is a local minimum in the middle. We can determine the location of this minimum by taking the derivative of Eq.~\eqref{amGR} with respect to $Y$ and equating it to zero. After some calculations, we obtain
\begin{equation}
\ln(a_{\textrm{min}})|_{\textrm{min}}=-\frac{1}{2}(1-2q_0+2\Omega_r)Y_{\textrm{min}},
\label{deY1}
\end{equation}
where $Y_\textrm{min}$ fulfils
\begin{eqnarray}
&&\ln[(1-j_0+2\Omega_r)Y_{\textrm{min}}]\nonumber\\
&=&\frac{-2+2(1-j_0)Y_{\textrm{min}}-3(1-2q_0)Y_{\textrm{min}}+\Omega_r Y_{\textrm{min}}}{2},\nonumber\\
\label{deY2}
\end{eqnarray}
in which the subscript $\textrm{min}$ denotes the local minimum. According to Eqs.~\eqref{deY1}, \eqref{deY2} and \eqref{alphaGR}, one can see that $Y_{\textrm{min}}$ increases as $j_0$ gets closer to $1+2\Omega_r$, with $\alpha$ getting closer to zero and $a_{\textrm{min}}$ approaching $0$ for a given $q_0$ in fit $(7)$. For example, we have found that the local minimum of $a_{\textrm{min}}\approx0.04$, corresponding to $\alpha\approx0.1$, $1-j_0+2\Omega_r=10^{-5}$ and $q_0=-0.561$; while the local minimum of $a_{\textrm{min}}\approx0.002$ corresponding to $\alpha\approx0.056$, $1-j_0+2\Omega_r=10^{-9}$ and $q_0=-0.561$. Note that the values of $j_0$ are compatible with fit $(7)$ in Ref.~\cite{Gruber:2013wua} and to get our above estimations we have simply set $\Omega_\kappa$ to zero.

According to these constraints and the asymptotic behaviours analyses in previous sections (the Universe would start from a finite past sudden singularity if $\alpha>2$ and a finite past type IV singularity if $0<\alpha\le2$), a universe based on the EiBI theory may start its expansion from a past type IV singularity with both $a_{\textrm{min}}$ and $\alpha$ being very small, as long as $j_0$ is very close to $1+2\Omega_r$, i.e., the radiation$+\Lambda$CDM model. It is, however, unlikely that the Universe starts from a sudden singularity on this case because this would require a relatively large value of $\alpha$ which would imply a too large value of $a_{\textrm{min}}$ which is incompatible with the history of the Universe.

\subsubsection{The analyses for negative Y in the EiBI theory}

\begin{figure}[!h]
\begin{center}
\includegraphics[scale=0.8]{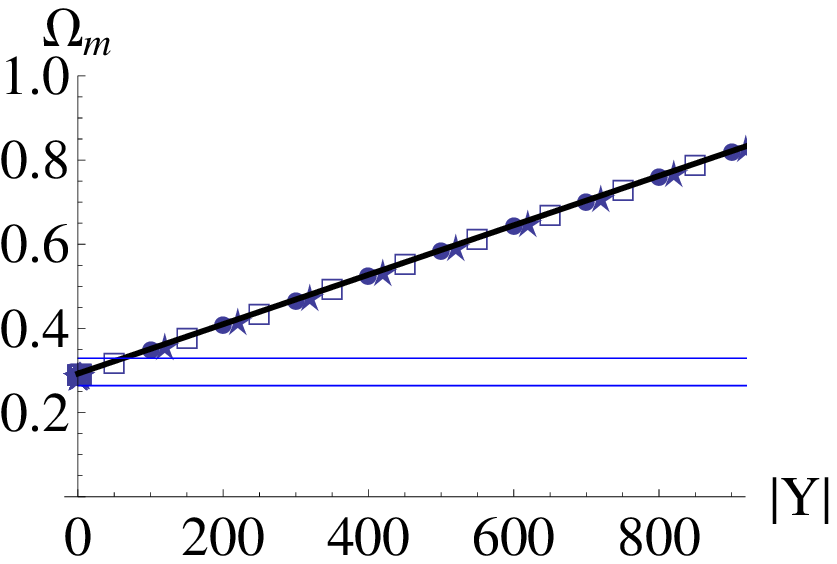}
\includegraphics[scale=0.8]{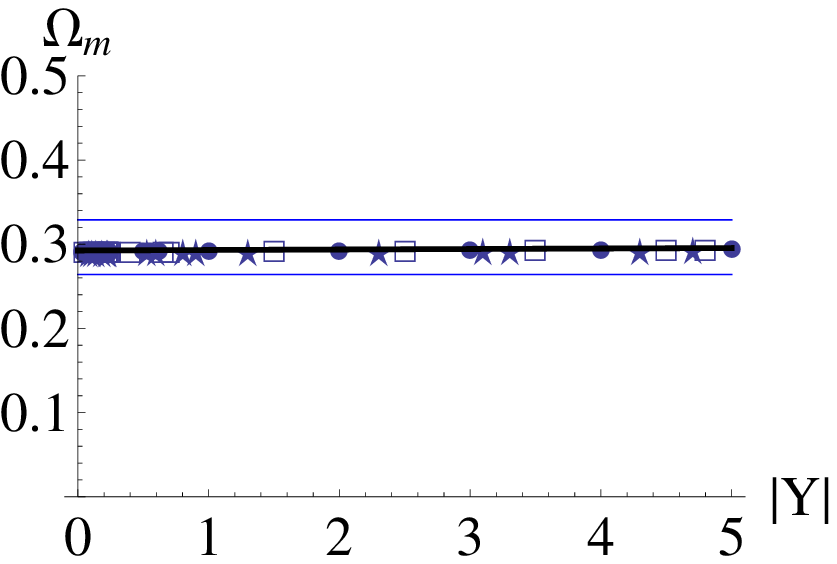}
\caption{The constraints on $\Omega_m$ from the assumptions $q_0=-0.561$, $j_0=1.001$ and $\Omega_r=8.48\times 10^{-5}$ for negative $Y$. On the top figure $|Y|$ changes from $0$ to $1/(j_0-1-2\Omega_r)$, and on the bottom one $|Y|$ changes from $0$ to $5$. The points in different symbols correspond to different $\Omega_\kappa$ with the same imposed
values in FIG.~\ref{om999}, and the black solid line represents the analytical result in GR given by Eq.~\eqref{omGR}. The region between the horizontal blue lines represents the $1\sigma$ errors of $\Omega_m$ derived from the given $q_0$ on the basis of the $\Lambda$CDM model.}
\label{om001}
\end{center}
\end{figure}

\begin{figure}[!b]
\begin{center}
\includegraphics[scale=0.8]{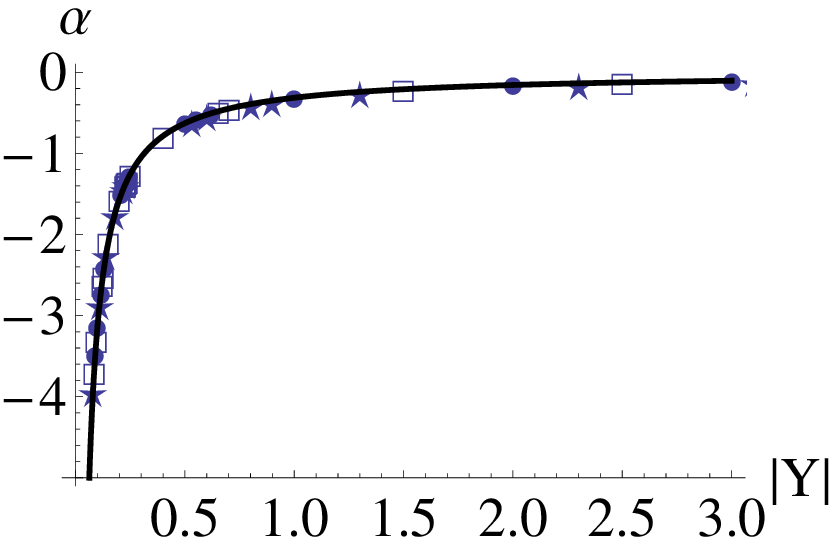}
\includegraphics[scale=0.8]{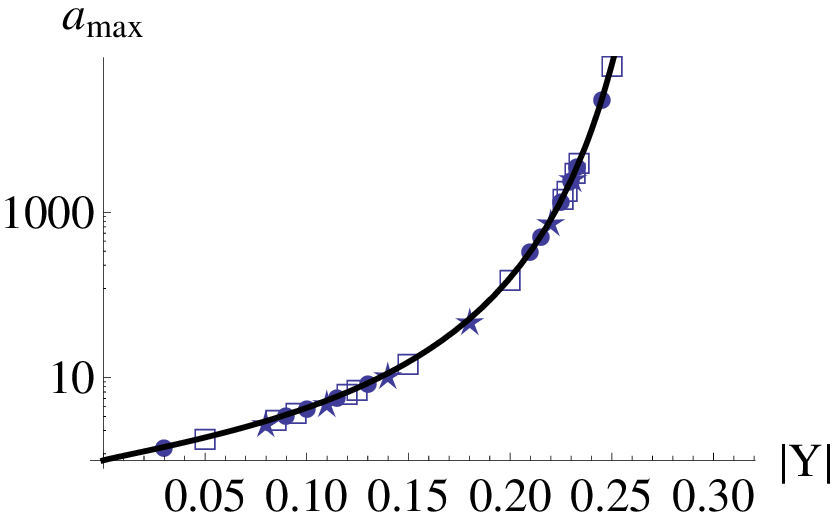}
\includegraphics[scale=0.8]{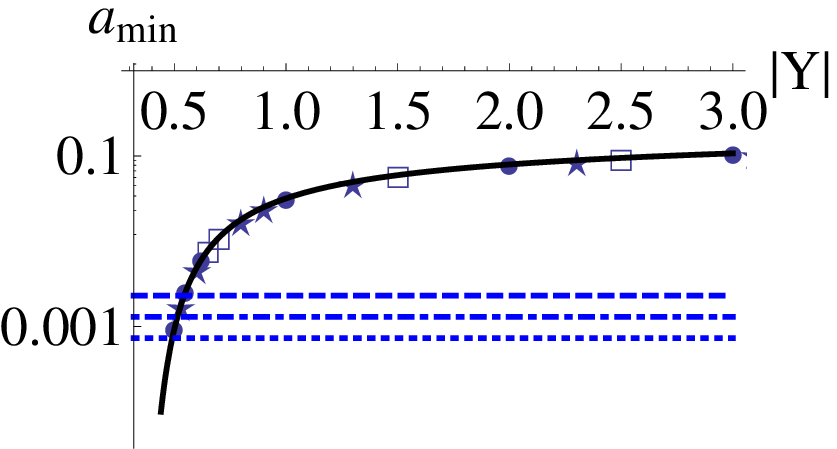}
\caption{The numerical results of $\alpha$ (top), $a_{\textrm{max}}$ (middle) for $\alpha<-1$, and $a_{\textrm{min}}$ (bottom) for $-1<\alpha<0$, versus $|Y|$. The points in different symbols denote the numerical results obtained on the basis of different $\Omega_\kappa$, and the black solid line represents the analytical results in GR. The smallest scale factor in which the loitering effect happens is shown as the blue horizontal lines ($\Omega_\kappa=10^{-6}$, $10^{-7}$, $10^{-8}$ from top to bottom).}
\label{alphaam001}
\end{center}
\end{figure}

In this subsubsection, we will carry out the analysis for negative $Y$, thus we have to assume that $1-j_0+2\Omega_r$ is negative according to Eq.~\eqref{defineY}. We make a prior assumption that $q_0$ and $j_0$ are independent for the fit and $j_0$ deviates in absolute value by the same amount from the $\Lambda$CDM model as in the model discussed on the previous subsubsection, that is, $q_0=-0.561$ and $j_0=1.001$. Note that with this assumption $j_0$ is within the $1\sigma$ errors of fit $(7)$ of Ref.~\cite{Gruber:2013wua}. One can see from FIG.~\ref{om001} that the numerical results of $\Omega_m$ for different $|Y|$ and a given $\Omega_\kappa$ are almost indistinguishable from the analytical GR result given in Eq.~\eqref{omGR} (Black line). Again, the two blue horizontal lines indicate the $1\sigma$ errors of $\Omega_m$, which is estimated from the assumed $q_0$ value based on the $\Lambda$CDM model. Furthermore, we obtain similar conclusions to those corresponding to the $Y$ positive case, analysed on the previous subsubsection, i.e.: (i) we do not see much difference between using EiBI and GR, the reason is that $\Omega_\kappa$ is very small as predicted in Refs.~\cite{Avelino:2012ge,Avelino:2012qe}. (ii) In order to obtain values of $\Omega_m$ compatible with the $\Lambda$CDM model, which we will consider as a guiding line of our analysis, we will stick to small values of $|Y|$ which we will consider to be small, roughly smaller than $5$ (see FIG.~\ref{om001}). For the sake of presenting our results in a clear and suitable way, we will highlight the region where $|Y|$ is smaller than $3$ (see FIG.~\ref{alphaam001}).

\begin{figure}[!h]
\begin{center}
\includegraphics[scale=0.8]{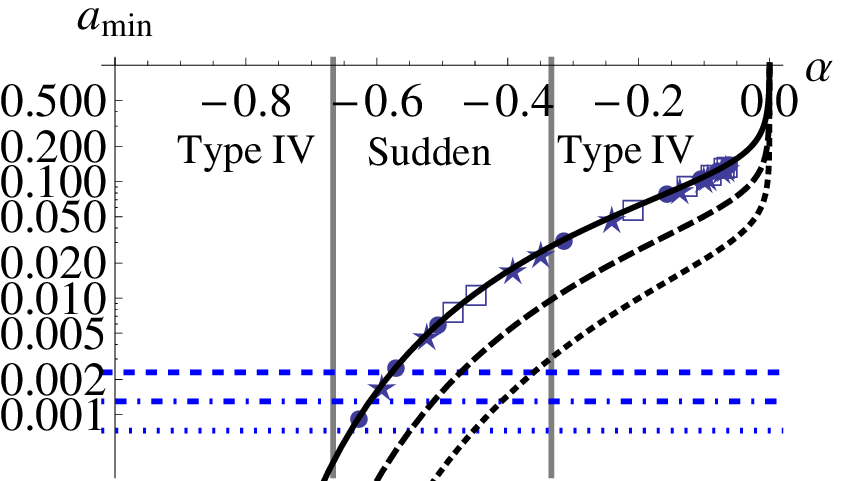}
\includegraphics[scale=0.8]{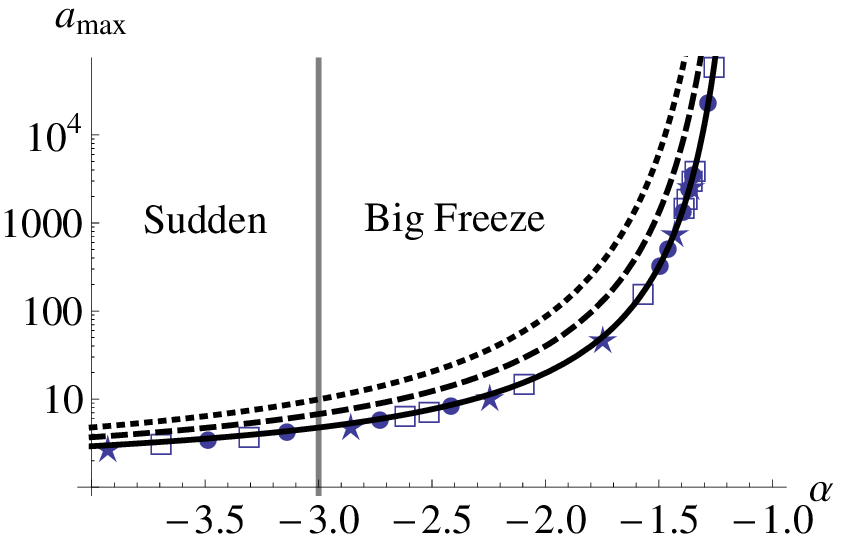}
\caption{These figures express $a_s$ in terms of $\alpha$. For the sake of convenience, we split the results into two different figures according to whether the singularities occur in the past (top) or in the future (bottom). The vertical grid lines splitting $\alpha$ classify different asymptotic singularities as proposed in TABLE.~\ref{summary}. Note that the quantised $\alpha$ cases in which the Hubble rate and its higher order derivatives are all regular are not shown in these figures. Again, the points in different symbols denote the numerical results in the EiBI theory with different $\Omega_\kappa$. Furthermore, the solid, dashed, and dotted curves indicate the analytical GR results combining Eqs.~\eqref{alphaGR} and \eqref{amGR} with $j_0-1-2\Omega_r=10^{-3}$, $10^{-4}$, and $10^{-5}$, respectively. The blue horizontal lines correspond to the scale factor in which the loitering effect happens in different $\Omega_\kappa$, with the same values chosen in FIG.~\ref{alphaam001}}
\label{amalpha}
\end{center}
\end{figure}

Similar to what we have done in the previous subsubsection, we can also numerically evaluate $\alpha$, $a_s$ for different $|Y|$ and $\Omega_\kappa$, then compare these results with the analytical results in Eqs.~\eqref{alphaGR} and \eqref{amGR} in GR. The results are shown in FIG.~\ref{alphaam001}. All these figures indicate again that the results in the EiBI theory are almost indistinguishable from those in GR for the $\Omega_\kappa$ values of interest \cite{Avelino:2012ge,Avelino:2012qe}. The first (top) figure shows that $\alpha\rightarrow0$ for large $|Y|$, while it approaches $-\infty$ for small $|Y|$. Note that because of the presence of radiation, there may be some particular scale factor $a_b>a_{\textrm{min}}$ in which the total pressure satisfies $\bar p=1$ thus implying a loitering effect at an infinite past. This particular scale factor value $a_b$ depends almost only on $\Omega_\kappa$ and its location is shown with the horizontal blue lines in FIG.~\ref{alphaam001}: lines from top to bottom correspond to decreasing $\Omega_\kappa$. One has to keep in mind that the value of $\alpha$ can determine what kind of singularity would happen in the Universe, no matter in GR or in the EiBI theory (see TABLE.~\ref{summary}). For the sake of convenience and completeness, we further show $a_s$ in terms of $\alpha$ in FIG.~\ref{amalpha}. In these figures, the points with different symbols again indicate different values of $\Omega_\kappa$ as imposed on the previous analyses. Furthermore, the vertical grid lines located on $\alpha=-1/3$, $-2/3$ on the top figure and $\alpha=-3$ on the bottom one classify different asymptotic singularities as summarized in TABLE.~{\ref{summary}}. Note that the quantised $\alpha$ cases in which the Hubble rate and its higher order derivatives are all regular are not shown in these figures. Moreover, the solid, dashed, and dotted curves indicate the analytical GR results combining Eqs.~\eqref{alphaGR} and \eqref{amGR} with $j_0-1-2\Omega_r=10^{-3}$, $10^{-4}$, and $10^{-5}$, respectively; and the blue horizontal lines correspond to the scale factor in which the loitering effect happens for different $\Omega_\kappa$, with the same values imposed in FIG.~\ref{alphaam001}.

The conclusions are the following: (i) If $\alpha<-1$, which implies the existence of future singularities, the values $\alpha<-3$ (sudden singularities), $\alpha=-3$ (type IV singularities) or $-3<\alpha<-1$ (big freeze singularities) are all compatible with the fit $(7)$ in \cite{Gruber:2013wua}. However, we cannot tell which of these singularities are preferred from an observational point of view and observational constraints on higher cosmographic parameters are necessary. It is worth mentioning that for a fixed $\alpha$, the closer to $1+2\Omega_r$ the jerk parameter at present $j_0$ is, the larger the maximum of the scale factor at the doomsday $a_{\textrm{max}}$ would be.

(ii) If $-1<\alpha<0$, the Universe will start either from a big loitering effect or a type IV singularity. A sudden singularity would not be allowed observationally because it will take place at a too large value of $a_{\textrm{min}}$ incompatible with the history of our universe (see FIG.~\ref{amalpha}). Moreover, it is also worth mentioning that if $j_0$ is getting much closer to $1+2\Omega_r$, that is, the radiation$+\Lambda$CDM model, the allowable region of $\alpha$ enlarges because $a_{\textrm{min}}$ decreases as $j_0$ gets closer to $1+2\Omega_r$ for a fixed $\alpha$. Finally, we evaluate $\Omega_{de}$, $\Omega_m$ $a_s$ and the dimensionless cosmic time between the singularities and the current time for various $\alpha$ and $\Omega_\kappa$ in TABLE.~\ref{big fit}. Note that the cases in which the past singularities are replaced with a big loitering effect are also shown.

\begin{table*}
  \begin{center}
    \begin{tabular}{||c|c|c|c|c|c||c|c|c|c|c|c||}
      \hline
      $\alpha$ &$\Omega_\kappa$ & $\Omega_m$ &$\Omega_{de}$ &$a_s$ (or $a_b$)&$H_0(t_s-t_0)$&$\alpha$ &$\Omega_\kappa$ & $\Omega_m$ &$\Omega_{de}$ &$a_s$ (or $a_b$)&$H_0(t_s-t_0)$ \\
      \hline
      $-3.5$ &$10^{-6}$&$0.292606$&$0.707309$&$3.5512$&$1.39474$&$-3$&$10^{-6}$&$0.292615$&$0.7073$&$4.7513$&$1.71593$\\
      &$10^{-7}$&$0.292606$&$0.707309$&$3.55106$&$1.39309$&&$10^{-7}$&$0.292615$&$0.7073$&$4.75106$&$1.71434 $\\
      &$10^{-8}$&$0.292606$&$0.707309$&$3.55104$&$1.3925$&&$10^{-8}$&$0.292615$&$0.7073$&$4.75103$&$1.71377$\\
      &$0$ (GR)&$0.292606$&$0.707309$&$3.55104$&$1.39217$&&$0$ (GR)&$0.292615$&$0.7073$&$4.75103$&$1.71345$\\
      \hline\hline
      $-2.5$ &$10^{-6}$&$0.292627$&$0.707288$&$7.67048$&$2.23619$&$-2$&$10^{-6}$&$0.292646$&$0.707269$&$19.7213$&$3.23296 $\\
      &$10^{-7}$&$0.292627$&$0.707288$&$7.66995$&$2.23468$&&$10^{-7}$&$0.292646$&$0.707269$&$19.7192$&$3.23158$\\
      &$10^{-8}$&$0.292627$&$0.707288$&$7.6699$&$2.23414$&&$10^{-8}$&$0.292646$&$0.707269$&$19.719$&$3.2311$\\
      &$0$ (GR)&$0.292627$&$0.707288$&$7.66989$&$2.23384$&&$0$ (GR)&$0.292646$&$0.707269$&$19.719$&$3.23083$\\
      \hline\hline
      $-0.9$ &$10^{-6}$&$0.292758$&$0.707157$&$0.00230578$&$-\infty$(Loitering)&$-0.8$&$10^{-6}$&$0.292784$&$0.707131$&$0.00230578$&$-\infty$(Loitering)\\
      &$10^{-7}$&$0.292759$&$0.707157$&$0.00129664$&$-\infty$(Loitering)&&$10^{-7}$&$0.292784$&$0.707131$&$0.00129664$&$-\infty$(Loitering)\\
      &$10^{-8}$&$0.292759$&$0.707157$&$0.0007$&$-\infty$(Loitering)&&$10^{-8}$&$0.292784$&$0.707131$&$0.000729153$&$-\infty$(Loitering)\\
      &$10^{-40}$&$0.292759$&$0.707157$&$7.3\times 10^{-12}$&$-\infty$(Loitering)&&$10^{-40}$&$0.292784$&$0.707131$&$1.54408\times 10^{-6}$&$-0.970345$\\
      &$10^{-43}$&$0.292759$&$0.707157$&$1.61021\times 10^{-12}$&$-0.970356$&&$10^{-43}$&$0.292784$&$0.707131$&$1.54408\times10^{-6}$&$-0.970345$\\
      &$0$ (GR)&$0.292759$&$0.707157$&$1.61021\times10^{-12}$&$-0.970356$&&$0$ (GR)&$0.292784$&$0.707131$&$1.54408\times10^{-6}$&$-0.970345$\\
      \hline\hline
      $-0.7$ &$10^{-6}$&$0.292817$&$0.707098$&$0.00230578$&$-\infty$(Loitering)&$0.15$&$10^{-6}$&$0.294282$&$0.705633$&$0.17496$&$-0.881185$\\
      &$10^{-7}$&$0.292817$&$0.707098$&$0.001$&$-\infty$(Loitering)&&$10^{-7}$&$0.294282$&$0.705633$&$0.174949$&$-0.881195$\\
      &$10^{-8}$&$0.292817$&$0.707098$&$0.0007$&$-\infty$(Loitering)&&$10^{-8}$&$0.294282$&$0.705633$&$0.174947$&$-0.881196$\\
      &$0$ (GR)&$0.292817$&$0.707098$&$1.5495\times10^{-4}$&$-0.970329$&&$0$ (GR)&$0.294282$&$0.705633$&$0.174947$&$-0.881196$\\
      \hline\hline
      $0.25$ &$10^{-6}$&$0.293592$&$0.706323$&$0.175573$&$-0.880798$&$0.2$&$10^{-6}$&$0.293851$&$0.706064$&$0.17372$&$-0.882178$\\
      &$10^{-7}$&$0.293592$&$0.706323$&$0.175562$&$-0.880806$&&$10^{-7}$&$0.293851$&$0.706064$&$0.17371$&$-0.882187$\\
      &$10^{-8}$&$0.293592$&$0.706323$&$0.175561$&$-0.880807$&&$10^{-8}$&$0.293851$&$0.706064$&$0.173709$&$-0.882188$\\
      &$0$ (GR)&$0.293592$&$0.706323$&$0.175561$&$-0.880807$&&$0$ (GR)&$0.293851$&$0.706064$&$0.173709$&$-0.882188$\\
      \hline
    \end{tabular}
    \caption{Using the first approach in Section~IV, here we show the constraints on the parameters derived for various $\alpha$ and $\Omega_\kappa$ based on fit $(1)$ of Ref.~\cite{Gruber:2013wua}. Here we assume $q_0=-0.561$, $\Omega_r=8.48\times 10^{-5}$, and $j_0=0.999$ or $1.001$, corresponding to positive or negative $\alpha$. The parameter constraints under the GR framework ($\Omega_\kappa=0$) are shown. Note that the cases in which the past singularities are replaced with a loitering effect in an infinite past are also shown. The additional analyses for $\Omega_\kappa=10^{-40}$ and $10^{-43}$ indicate that as long as $\Omega_\kappa$ is small enough, and $\alpha$ is close enough to $-1$, it is possible to derive a small enough $a_s$ to stand for the existence of a past type IV singularity in the EiBI theory. We have to stress that the allowable region of $\alpha$ in which a small enough $a_s$ can be obtained also increases as $j_0$ gets close to $1+2\Omega_r$, as mentioned in this section.}
    \label{big fit}
  \end{center}
\end{table*}

\subsection{The second approach: assuming $\Omega_m$}

In the previous subsection, we only consider fit $(7)$, which is the closest fit in Ref.~\cite{Gruber:2013wua} to the radiation$+\Lambda$CDM model. Additionally, we can also use other fits to constrain our model. One can see that other fits from $(1)$ to $(6)$ have a significant difference from fit $(7)$: The jerk parameter $j_0$ is different from $1+2\Omega_r$ by a comparable amount. This fact makes these data sets deviate a lot from the radiation$+\Lambda$CDM model and when applied to the model we are analysing we get a too large $a_\textrm{min}$ which is incompatible with the history of the Universe, as we mentioned previously. Therefore, we will only analyse the models in which future singularities happen with the data sets from fits $(1)$ to $(6)$.

To analyse the future singularities with these data sets, we use another approach different from the one we followed on the previous subsection: we fix the value of $\Omega_m$ according to the Planck mission \cite{CMB} and assume it to be model independent, then we assume that $\Omega_\kappa$ is roughly within the range $10^{-7}$ to $10^{-4}$. We find that only fits $(2)$ and $(3)$ are compatible with the analyses of the cases in which $\alpha<-1$. The reason is that for the cases in which $\alpha<-1$, the derivative of $\bar p_{de}/\bar\rho_{de}$ with respect to the scale factor should be negative. Furthermore, the value of $\bar p_{de}/\bar\rho_{de}$ should be smaller than $-1$. On the basis of GR, these criteria are only valid in these two data sets fits $(2)$ and $(3)$. Interestingly, the authors of \cite{Gruber:2013wua} also claimed that these two fits, in addition to fit $(7)$, are the most reasonable results of their analyses. Hence, we use the data in fits $(2)$ and $(3)$, set the values of $\Omega_m=0.315$, $\Omega_r=8.48\times 10^{-5}$, and $\Omega_\kappa$ from $10^{-7}$ to $10^{-4}$, and numerically solve the resulting $\alpha$, $\Omega_{de}$, $a_{\textrm{max}}$ as well as the dimensionless cosmic time elapsed from the current time to the doomsday $H_0(t_\textrm{max}-t_0)$. The results are shown in TABLE~\ref{fit 2} and TABLE~\ref{fit 3}.

According to TABLE~\ref{fit 2} and TABLE~\ref{fit 3}, one can see that fit $(2)$ prefers the parameter space $-3<\alpha<-1$, implying a big freeze singularity for the death of the Universe, while fit $(3)$ prefers the parameter space $\alpha<-3$ where the occurrence of a sudden singularity is preferred.

\begin{table}[!h]
  \begin{center}
    \begin{tabular}{||c||c|c|c|c||}
      \hline
      $\Omega_\kappa$ & $\alpha$ &$\Omega_{de}$ &$a_\textrm{max}$&$H_0(t_\textrm{max}-t_0)$ \\
      \hline
      $0$ (GR)&$-1.94103$&$0.684915$&$2.05115$&$0.621529$\\
      $10^{-7}$&$-1.94103$&$0.684915$&$2.05115$&$0.622249$\\
      $10^{-6}$&$-1.94103$&$0.684916$&$2.05115$&$0.6235$\\
      $10^{-5}$&$-1.94102$&$0.684919$&$2.05113$&$0.62681$\\
      $10^{-4}$&$-1.94095$&$0.684948$&$2.05102$&$0.635257$\\
      \hline
    \end{tabular}
    \caption{The constraints of the parameters derived according to the data fit (2) in Ref.~\cite{Gruber:2013wua} where $H_0=70.25$, $q_0=-0.683$, $j_0=2.044$. Here we use the second approach presented in Section~IV in which we assume $\Omega_m=0.315$ according to the Planck data and $\Omega_\kappa=10^{-4}$, $10^{-5}$, $10^{-6}$, $10^{-7}$. The parameter constraints under the GR framework ($\Omega_\kappa=0$) are also shown.}
  \label{fit 2}
  \end{center}
\end{table}

\begin{table}[!h]
  \begin{center}
    \begin{tabular}{||c||c|c|c|c||}
      \hline
      $\Omega_\kappa$ & $\alpha$ &$\Omega_{de}$ &$a_\textrm{max}$&$H_0(t_\textrm{max}-t_0)$ \\
      \hline
      $0$ (GR)&$-3.19514$&$0.684915$&$1.39279$&$0.317249$\\
      $10^{-7}$&$-3.19514$&$0.684915$&$1.39279$&$0.318152$\\
      $10^{-6}$&$-3.19514$&$0.684916$&$1.39279$&$0.31972$\\
      $10^{-5}$&$-3.19516$&$0.68492$&$1.39276$&$0.323837$\\
      $10^{-4}$&$-3.19532$&$0.68496$&$1.39248$&$0.334114$\\
      \hline
    \end{tabular}
    \caption{The constraints on the model parameters derived according to the data fit (3) in Ref.~\cite{Gruber:2013wua} where $H_0=70.09$, $q_0=-0.658$, $j_0=2.412$. Here we use the second approach presented in Section~IV in which we assume $\Omega_m=0.315$ according to the Planck data and $\Omega_\kappa=10^{-4}$, $10^{-5}$, $10^{-6}$, $10^{-7}$. The parameter constraints under the GR framework ($\Omega_\kappa=0$) are also shown.}
  \label{fit 3}
  \end{center}
\end{table}

\section{conclusions}
The Eddington-inspired-Born-Infeld theory (EiBI) proposed recently is characterised by being equivalent to Einstein theory in vacuum but differing from it in the presence of matter. Most importantly, it also features the ability to avoid some singularities such as the Big Bang singularity in the finite past of the Universe, and the singularity formed after the collapse of a star. It is hence interesting to see whether this ability to avoid/smooth other kinds of singularities, especially, those driven by the phantom dark energy which could be responsible for the current accelerating expansion of the Universe, is efficient enough or not. 

In this paper, we give a thorough analysis on the avoidance of all dark energy related singularities by deriving the asymptotic behaviours of the Hubble rate and the cosmic time derivatives of the Hubble rate defined by the physical metric $g_{\mu\nu}$ coupled to matter, and by the auxiliary metric $q_{\mu\nu}$ compatible with the physical connection. For the physical metric $g_{\mu\nu}$ we find that though the Big Rip singularity and the Little Rip event driven by phantom dark energy are not cured in the EiBI theory, this theory to some extent smooth the other phantom dark energy related singularities present in GR by leaving some region of the parameter space in which the future Big Freeze singularity is altered into a future sudden or future type IV singularity. Additionally, the past  singularity present in GR is also smoothed in this theory in some parameter space as a past type IV singularity. Note that a past type IV singularity present in GR is, in some parameter space, worsened into a past sudden singularity while smoothed as a regular birth of the Universe at some quantized parameter space or even as a loitering effect in an infinite past. As for the auxiliary metric $q_{\mu\nu}$ compatible with the physical connection (we remind that the EiBI setup we are dealing with is formulated \textit{\`{a} la} Palatini formalism), all the dark energy related singularities of interest are avoided except for some very specific parameter space in which the past Type IV singularities of the auxiliary metric still exist (See TABLE.~\ref{summary} for a summary).

Furthermore, we analysed the fate of a bound structure near the singularities of the EiBI theory. We find that the bound structure would be destroyed before the Universe approaches a big rip singularity and a little rip event, while remains bounded at a sudden, big freeze and type IV singularities.

Besides, we also use the cosmographic approach, which is characterised by its theoretical model-independence, to constrain the parameters present in our model, so that we in principle can forecast the doomsdays and describe the birth of the Universe based on our model. As a result, it turns out that the cosmographic analyses pick up the physical region which determines the occurrence of a type IV singularity in the finite past or the loitering effect in an infinite past. While it is necessary to impose more conditions, such as the use of higher order cosmographic parameters with more accurate observations or other physical constraints, to forecast the future doomsdays of the Universe in this model. According to these results, the EiBI theory is indeed a reliable theory which is able to cure or smooth the singularities predicted originally in GR, thus it makes the theory a convincing alternative to GR as a way to smooth singularities. 

\acknowledgments

M.B.L. is supported by the Basque Foundation for Science IKERBASQUE.
She also wishes to acknowledge the hospitality of LeCosPA Center at the National Taiwan University during the completion of part of this work and the support of the Portuguese Agency ``Funda\c{c}\~{a}o para a Ci\^{e}ncia e Tecnologia" through PTDC/FIS/111032/2009 and partially by the Basque government Grant No. IT592-13.
C.-Y.C. and P.C. are supported by Taiwan National Science Council under Project No. NSC 97-2112-M-002-026-MY3 and by Taiwan’s National Center for Theoretical Sciences (NCTS). P.C. is in addition supported by US Department of Energy under Contract No. DE-AC03-76SF00515.

\end{document}